\documentclass{article}

\usepackage{PRIMEarxiv}
\usepackage{amsmath,multirow}
\usepackage{array,tabularx}
\usepackage[ruled]{algorithm2e}
\usepackage[utf8]{inputenc} 
\usepackage[T1]{fontenc}    
\usepackage{hyperref}       
\usepackage{url}            
\usepackage{booktabs}       
\usepackage{amsfonts}       
\usepackage{nicefrac}       
\usepackage{microtype}      
\usepackage{lipsum}
\usepackage{fancyhdr}       
\usepackage{graphicx, subfigure} 
\graphicspath{{media/}}     

\title{Graph Neural Networks with Motif-aware for Tenuous Subgraph Finding
\thanks{\textit{\underline{Citation}}: 
\textbf{Heli sun, Miaomiao Sun, Liang He, and Xiaolin Jia. Graph Neural Networks with Motif-aware for Tenuous Subgraph Finding. 15 Pages.... DOI:https://doi.org/10.1145/nnnnnnn.nnnnnnn.}} 
}

\author{
  Heli sun \\
  Xi'an Jiaotong University \\
  Xi'an \\
  China\\
  \texttt{\{Heli sun \}hlsun@mail.xjtu.edu.cn} \\
   \And
  Miaomiao Sun \\
  Xi'an Jiaotong University \\
  Xi'an \\
  China\\
  \texttt{miaomiaosun@stu.xjtu.edu.cn} \\
   \AND
   Liang He \\
   Xi'an Jiaotong University \\
   Xi'an \\
  China\\
   \texttt{lh@mail.xjtu.edu.cn} \\
   \And
   Xiaolin Jia \\
   Xi'an Jiaotong University \\
   Xi'an \\
  China\\
   \texttt{xlinjia@mail.xjtu.edu.cn} \\
}

\begin{document}
\maketitle

\begin{abstract}
Tenuous subgraph finding aims to detect a subgraph with few social interactions and weak relationships among nodes. Despite significant efforts have been made on this task, they are mostly carried out in view of graph-structured data. These methods depend on calculating the shortest path and need to enumerate all the paths between nodes, which suffer the combinatorial explosion. Moreover, they all lack the integration of neighborhood information. To this end, we propose a novel model named Graph Neural Network with Motif-aware for tenuous subgraph finding (GNNM), a neighborhood aggregation based GNNs framework which can capture the latent relationship between nodes. Specially, we design a GNN module to project nodes into a low dimensional vector combining the higher-order correlation within nodes based on a motif-aware module. And then design greedy algorithms in vector space to obtain a tenuous subgraph whose size is greater than a specified constraint. Particularly, considering existing evaluation indicators cannot capture the latent friendship between nodes, we introduce a novel Potential Friend($PF$) concept to measure the tenuity of a graph from a new perspective. Experimental results on the real-world and synthetic datasets demonstrate that our proposed method GNNM outperforms existing algorithms in efficiency and subgraph quality.
\end{abstract}


\section{Introduction}

Recently, researchers have focused on the problem of tenuous subgraph finding. Different from dense groups finding\cite{YangWSDM2013,Xie2011,huang2014,BothorelCMM15,sanei2018,conte2018,Wen2019,Liu2020,U2001}, tenuous subgraph finding aims to obtain a subgraph with few social interactions and weak relationships among nodes. It has a wide range of applications and one of the important tasks is reviewer selection. Conference program chairs need to assign experts to review papers. Besides matching the expertise of reviewers with the topics of submissions, it is crucial to avoid assigning reviewers socially close to each other and the authors of the paper to ensure unbiased assessments. The problems of psycho-educational group formation and friend recommendation in weak social networks are also related to tenuous subgraph finding.

\begin{figure}[htbp]
	\centerline{\includegraphics[scale=0.65]{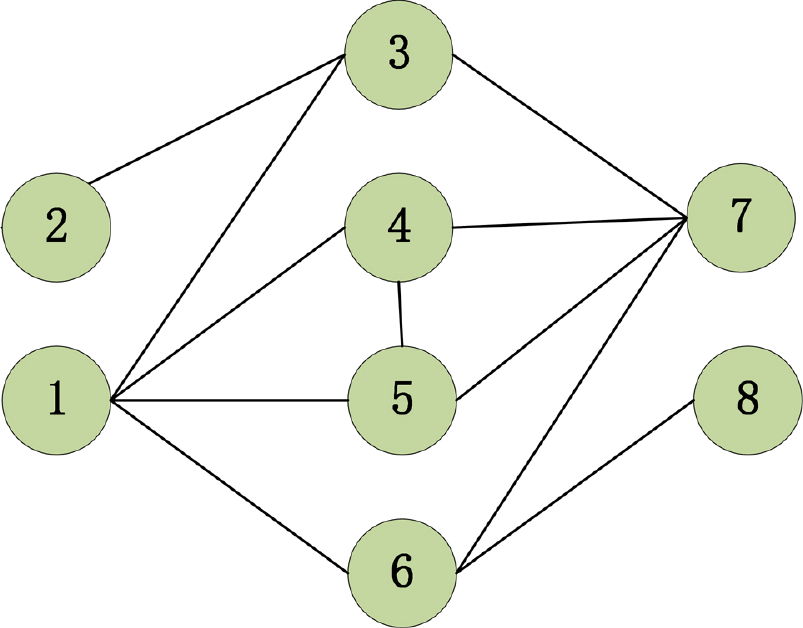}}
	\caption{A toy example of social network}
	\label{fig:a part of social network}
\end{figure}

Some works have focused on the problem of tenuous subgraph finding. Shen\cite{Shen2020} et al. modeled the problem of tenuous subgraph finding as Minimum $k$-Triangle Disconnected Group(MkTG) which proposes the concept of $k$-triangle to measure the tenuity of a subgraph. $K$-triangle is defined as a triple in a subgraph, in which the shortest path between any two nodes is less than $k$. The objective of MkTG is to get a subgraph in which the number of $k$-triangles is minimized. Li\cite{Li2018} et al. then formulated this problem as $K$-Line-Minimized(KLM) which presents $k$-line to measure the tenuity of a subgraph. $K$-line is defined as a pair of nodes in a subgraph whose shortest path is less than $k$. KLM aims at obtaining a subgraph in which the number of $k$-lines is minimized. Although these methods could solve the problem of tenuous subgraph finding, there remain some concerns. When measuring the shortest path between two nodes, they need to enumerate all the paths between nodes, which brings about the combinatorial explosion problem. Thereby, these methods suffer high time cost when computing the number of nodes involved in $k$-triangles or $k$-lines. Moreover, they lack the integration of neighborhood information and may do not work well in some special cases. More concretely, as shown in Fig. \ref{fig:a part of social network}, we may get a tenuous subset $T \in\{1,2,7\}$ because the nodes in it do not form any 1-line or 1-triangle. Nevertheless, it is not an optimal subset to form a tenuous subgraph because these nodes have many common neighbors, which indicates that they may have frequent interaction and be familiar with each other. And we argue that another tenuous subset $T \in\{2,7,8\}$ may be a better choice.

Meanwhile, the weights between connected nodes should be adjusted based on the network motifs\cite{Milo824} they participate in. Network motif, also known as the higher-order structure of a network, refers to a small sub-network structure. As shown in Fig. \ref{fig:a part of social network}, node 7 has 3, 4, 5, and 6 four neighbors, and nodes 4, 5, and 7 form a 3-node motif. If only considering the neighbor information of nodes, we could conclude that these nodes have the same influence on node 7. However, the existence of the 3-node motif reveals that the relationship between the three nodes 4, 5, and 7 is as close as an iron triangle. Consequently, nodes 4, 5 have greater influences on node 7.  

Motivated by the aforementioned observations, we introduce a new model GNNM by taking the above facts into tenuous subgraph finding. Specially, to integrate high-order structures within the neighborhoods, we first design a motif-aware module to calculate the higher-order correlation between nodes based on 3-node motifs and assign weights to different connected nodes. Then we apply one kind of neighborhood aggregation based GNNs to explicitly incorporate such correlation into the neighborhood aggregation which maps the original network to a low-dimensional vector space. Finally, a heuristic algorithm is exploited to get a subgraph whose size is greater than a specified constraint. In terms of evaluation indicators, considering $k$-line or $k$-triangle is limited in some cases, we propose a new concept $PF$ to measure the tenuity of a subgroup. Experimental results demonstrate that the proposed method GNNM outperforms existing algorithms in efficiency and subgraph quality. 
The main contributions of this paper are summarized as follows. 
\begin{itemize}
	\item We propose a novel model for tenuous subgraph finding problem utilizing motif-based network representation. To the best of our knowledge, it is the first attempt to address tenuous subgraph finding problem in the vector space.

	\item We develop a GNN module combining higher-order structural information obtained by the motif-aware module that extracts local features from a node’s neighbors with different weights.

	\item We introduce a new indicator $PF$ to measure the tenuity of a subgraph which unifies the indicators of $k$-line and $k$-triangle and combines both advantages of them, which can discover the latent friendship between nodes.

	\item We conduct extensive experiments to validate the efficiency and effectiveness of our method. We also verify the effectiveness of motif-based network representation.
\end{itemize}
The rest of this paper is organized as follows. Section 2 briefly introduces related work. Section 3 proposes the problem definition. Section 4 presents the framework of our proposed model in detail. In Section 5, we show the experimental results. Finally, Section 6 concludes the paper.

\section{Related Work}

\subsection{Tenuous Subgraph Finding}

Tenuous subgraphs have many practical significances, e.g. reviewer selection and psychoeducational group formation. Watrigant\cite{Watrigant} et al. studied the problem of ﬁnding a subgraph of $k$ nodes with the fewest induced edges on the chordal graph. Shen\cite{shen2015} et al. proposed a method for ﬁnding a group in which individuals are not familiar with each other. Later, Shen\cite{Shen2020} et al. proposed the MkTG problem for finding tenuous subgraphs which uses the number of $k$-triangles in the subgraph to measure the tenuity of the subgraph and designs a TERA algorithm to solve the problem. The MkTG problem aims to find a subgraph in which the number of $k$-triangles is as few as possible, while the number of nodes is as many as possible. A $k$-triangle is a triplet of nodes $\{\mathrm{u}, \mathrm{v}, \mathrm{w}\} \in \mathrm{G}$, such that $d_{G}(u, v) \leq k$, $d_{G}(u, w) \leq k$, $d_{G}(v, w) \leq k$, where $k$ can be any positive integer, $d_{G}(u, v)$ is the shortest distance between $(u,v)$ in graph $G$. Although TERA can find tenuous subgraphs, as mentioned before, $k$-triangle is limited in some cases. What is more, it needs to calculate the number of $k$-triangles that each node participates in which leads to high computation. Based on the MkTG problem, Li\cite{Li2018} et al. then proposed the KLM problem for tenuous subgraph finding which uses $k$-line to measure the tenuity of a subgraph and designs several algorithms to solve the problem. $K$-line is a node pair $\{\mathrm{u}, \mathrm{v}\} \in \mathrm{G}$, where $d_{G}(u, v) \leq k$. The KLM problem aims to find a subgraph in which the number of $k$-lines is as few as possible, while the number of nodes is as many as possible. However, $k$-line is also limited in some cases. Different from the previous work, Hsu \cite{hsu2018} et al. proposed the UTNA problem using tenuous subgraph finding to solve the problem of group therapy. Group therapy is one of the main clinical methods for the treatment of mental illness, but the formation of the treatment team is quite challenging. UTNA aims to automatically find the largest and reasonable subgraph so that the number of connected edges is as few as possible, and nodes satisfy Noah’s Ark Principle. Li \cite{li2020} studied the problem of finding tenuous subgraphs in attributed networks that contain specific vertices.

The above-mentioned methods are all based on the graph-structured data to find tenuous subgraphs. They all propose different definitions of tenuous subgraphs, and then design an algorithm based on these definitions. Different from previous methods, in this paper, we solve the problem of tenuous subgraph finding by mapping the original graph-structured data into the low-dimensional vector.

\subsection{Network Representation}

The problem of network representation\cite{Cai2017,Zhang2020} aims to map each node $v \in V$ into a low-dimensional vector, i.e., learning a function $f: \mathrm{v} \rightarrow R^{h}$, where $h$ is is the dimension of vector and usually far less than $|V|$. By mapping each node in the network as a low-dimensional dense vector, the topological structure and feature information of the original network can be efficiently stored in the learned latent vector. 

DeepWalk\cite{deepwalk} performs a fixed-length random walk and then uses the Skip-Gram model to learn the representation vector of nodes. Node2Vec\cite{node2vec} designs a biased random walk method based on Deepwalk, which is a trade-off between depth-first and breadth-first search. LINE\cite{line} tries to preserve the first-order and second-order similarity in the network. GraRep\cite{grarep} first obtains a co-occurrence matrix, and then uses SVD to obtain the low-dimensional vectors. TADW\cite{tadw} is a network representation method for text attribute networks. 

More recently, success in extending deep learning to graphs brought about Graph Neural Networks (GNNs)\cite{Wu2021,Zhou2021}. Being powerful yet efficient, neighborhood aggregation based GNNs has hence attracted the attention of numerous research works. The majority of GNNs learn the representation vectors of nodes via aggregating and transforming the features within their neighborhoods. Graph autoencoders are one of the methods. GAE\cite{gae} is a kind of variational graph autoencoder that replaces the linear layer with the graph convolutional layer. Structural Deep Network Embedding(SDNE)\cite{sdne} maintains the proximity between 2-hop neighbors through a deep autoencoder. Deep Neural Network for Graph Representations(DNGR)\cite{dngr} uses a random surfing strategy to capture graph structure information, and then trains stacked denoising autoencoders to embed nodes. 

In this paper, we mainly use the graph autoencoder based on graph convolutional network\cite{kipf2017,gao2018} to learn the node representations from graph topological structure and node features information.

\subsection{Network Motif}

There may be some important high-order structures in the network, which will affect the correlation between nodes. We call these structures network motifs. Network motifs are building blocks in complex networks and some works have proved that motifs are essential to the understanding of higher-order structural information in networks. For example, the existence of a triangle may indicate that the relationship between these three nodes in a social network is as close as a iron triangle. Researchers have shown great interests in the field of network motif recognition technology. The recognition of motifs in the network has attracted more and more attention, and it has been widely used in social networks\cite{Ugander2013SubgraphFM,Rahmtin2017}, biology networks\cite{bioinformatics}, and so on. Most of the existing works mainly focus on how to count motif subgraphs\cite{Ahmed2015,HanS16,Pinar2017}. Recently, it has been proven that motifs can also be used for graph clustering or community detection\cite{Benson163,Yin2017} tasks. And some works propose to use different motif patterns to discover different topics in the network\cite{Long2020}. 

Compared with previous studies, we explore the high-order structures based on 3-node motifs, and then fuse this high-order relationship into network representation to extract local features from a node’s neighbors with different weights. To the best of our knowledge, it is the first attempt to apply network motif in the tenuous subgraph finding problem to avoid introducing more potential relationship between nodes.

\section{Problem Definition}

In this section, we first give some definitions. Then we present the formal definition of tenuous subgraph finding. The notations in this paper are summarized in Table \ref{tab:symbols}.
\begin{table}[]
\centering
\caption{Notations}\label{tab:symbols}
\setlength{\tabcolsep}{20mm}{
\begin{tabular}{ll}
\hline
Symbols & Descriptions                                  \\ 
\hline
A       & graph adjacency matrix                        \\ 
X       & node features matrix                          \\ 
N       & number of nodes in graph                      \\ 
M       & co-occurrence matrix                          \\ 
Z       & node representations                          \\ 
T       & tenuous subset                                \\ 
h       & hidden dimension                              \\ 
D       & number of node features                       \\ 
A'      & reconstructed adjacency matrix                \\ 
D'      & diagonal degree matrix                        \\ 
S       & set for defining motif-based adjacency matrix \\ 
K       & size of tenuous subgraph                      \\ 
\hline
\end{tabular}}
\end{table}

To find tenuous subgraphs, we need to define the tenuity of a subgraph first. Considering existing evaluation indicators cannot capture the latent friendship between nodes, we introduce a novel notion $PF$ for measuring the tenuity of subgraphs.

Given an undirected, unweighted graph $G=(V, E)$. Let $N_{k}(u)$ denote the set of nodes at the shortest distance exactly no more than $k$ from $u$ in $G$. Note that $N_{1}(u)$ denotes the set of neighbors of node $u$.

\textbf{Definition 1.} Common Neighbors ($CN_{k}$)

For any two nodes $(u,v) \in E$, the $k$ hop common neighbors of them are defined as follows:
\begin{equation}
C N_{k}(u, v)=\{(u, v, w \mid w \in (N_{k}(u) \cap N_{k}(v)))\}
\end{equation}

Note that common neighbors of two nodes used here contain the nodes themselves.

\textbf{Definition 2.} Potential Friends ($PF$)

Given an undirected, unweighted subgraph $T=(V, E)$, for any two nodes $u$ in $T$, $v$ in $T$, $P F_{T}$ are defined as follows:

\begin{equation}
P F_{T}=\left\{\begin{array}{lr}
\left\{C N_{k}(u, v)\right\} & \text { if } v \notin N_{k}(u) \\
\left\{C N_{k}(u, v) \cup(u, v)\right\} & \text { otherwise }
\end{array}\right.
\end{equation}

We use $PF$ to measure the tenuity of a graph based on above definition. This measurement can be seen as an augment of $k$-line and $k$-triangle. By this measurement, we can not only combine $k$-line and $k$-triangle, but also capture the potential friends in the obtained subgraph.

Let $L$ denote the set of node pairs that form $k$-lines, and $R$ denote the set of triplets in subgraph $T$ that form $k$-triangles respectively. Suppose $T_{1}$ represents the set of node pairs whose shortest distance are no more than $k$, and $T_{2}$ represents the set of rest node pairs in $T$. $L_{T}$ and $R_{T}$ are defined as follows:

\begin{equation}
L_{T}=\{(u, v) \mid (u,v)\ forms\ a\ k-line\}
\end{equation}
\begin{equation}
R_{T}=\{(u, v, w) \mid (u,v,w)\ forms\ a\ k-triangle\}
\end{equation}

\textbf{Theorem 1.} $PF_{T}$ contains $L_{T}$, i.e. $L_{T} \subseteq PF_{T}$.

$Proof.$  For convenience, $P F_{T}$ can be presented as $P F_{T}=P F_{T 1} \cup P F_{T 2}=\left\{C N_{k}(u, v) \cup(u, v) \mid u \in T_{1}, v \in T_{1}\right\} \cup \left\{C N_{k}(u, v) \mid u \in T_{2}, v \in T_{2}\right\}$. Due to $T_{2}$ represent the set of node pairs whose shortest distance are more than $k$ in $T$, thus $L_{T_{2}}= \emptyset$. $L_{T}$ can be presented as $L_{T}=L_{T_{1}} \cup L_{T_{2}}=L_{T_{1}}=\left\{(u, v) \mid u \in T_{1}, v \in T_{1}\right\}$. $P F_{T} \cap L_{T}=(P F_{T_{1}} \cap L_{T_{1}}) \cup (P F_{T_{2}} \cap L_{T_{2}})=P F_{T_{1}} \cap L_{T_{1}}$.

For any node pair $(u, v) \in T_{1}$, we can find that $L_{T_{1}}= \{(u, v) \mid u \in T_{1}, v \in T_{1}\}$ from formula (3). Since $P F_{T 1}=\left\{C N_{k}(u, v) \cup(u, v) \mid u \in T_{1}, v \in T_{1}\right\}$. Thus $P F_{T_{1}}$ contains $L_{T_{1}}$, $P F_{T_{1}} \cap L_{T_{1}}= \{(u, v) \mid u \in T_{1}, v \in T_{1}\}$. In general, $P F_{T} \cap L_{T}= \{(u, v) \mid u \in T_{1}, v \in T_{1}\}$. Therefore, we can come to a conclusion that $PF_{T}$ contains $L_{T}$.

\textbf{Theorem 2.} $PF_{T}$ contains $R_{T}$, i.e. $R_{T} \subseteq P F_{T}$.

$Proof.$ Since nodes in $T_{2}$ cannot form any $k$-triangles, thus $R_{T_{2}}= \emptyset$. $R_{T}$ can be represented as $R_{T}=R_{T_{1}} \cup R_{T_{2}}=R_{T_{1}}=\{(u, v, w) \mid u \in T_{1}, v \in T_{1},w \in T_{1}\}$. $P F_{T} \cap R_{T}=(P F_{T_{1}} \cap R_{T_{1}})\cup (P F_{T_{2}} \cap R_{T_{2}})=P F_{T_{1}} \cap R_{T_{1}}$.

For any node pair $(u, v) \in T_{1}$, first, the shortest distance between $(u,v)$ are no more than $k$; second, $C N_{k}(u, v)$ are the set of nodes whose shortest distance between $u$ and $v$ are no more than $k$, that is they form a $k$-triangle. Thus $P F_{T_{1}} \cap R_{T_{1}}=\left\{C N_{k}(u, v) \mid u \in T_{1}, v \in T_{1}\right\}$. In general, $P F_{T} \cap R_{T}=\left\{C N_{k}(u, v) \mid u \in T_{1}, v \in T_{1}\right\}$. Therefore, we can come to a conclusion that $PF_{T}$ contains $R_{T}$.

Based on the above definitions, we give the formal definition of tenuous subgraph finding problem which obtains a subgraph by minimizing the number of $|PF|$.

\textbf{Problem: Tenuous subgraph finding}

Given an undirected, unweighted attribute graph $G = (V, E, X)$, size constraint $K$, returns a tenuous subset $T$, where $T$ satisfies the following conditions:
\begin{itemize}
	\item $|PF_{T}|$ is minimized
	\item $|T| \geq K$
\end{itemize}
where $V=\left\{v_{1}, v_{2}, \cdots, v_{n}\right\}$ is the vertex set with $n$ nodes in total, $E$ is the edge set,  $X=\left(x_{1}, x_{2}, \cdots, x_{n}\right)^{D}$ is the feature matrix, and $|T|$ is the number of nodes in $T$.

\begin{figure}[htbp]
	\centerline{\includegraphics[scale=0.3]{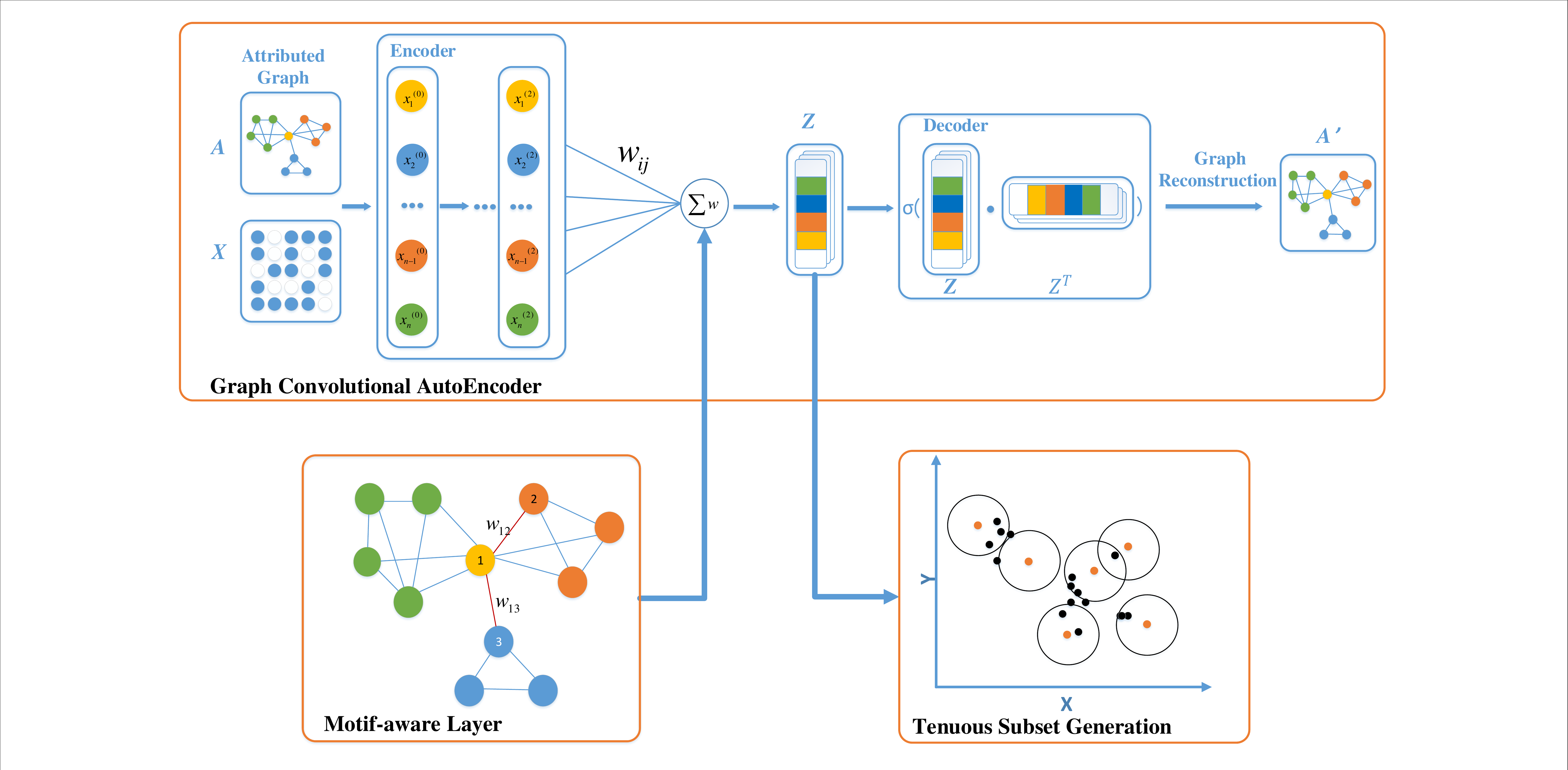}} 
	\caption{The framework of GNNM. Left-bottom is the motif-aware module. Above is the graph convolutional autoencoder module which learns node representations from graph structure and node features based on motif-aware module. Right-bottom is the tenuous subset generation module which is used to detect a node set that meets certain conditions.}
	\label{fig:framework}
\end{figure}

\section{Proposed Model}

In this section, we introduce the framework of proposed model GNNM. As shown in Fig. \ref{fig:framework}, there are three modules in our framework: motif-aware module, graph convolutional autoencoder module, and tenuous subset generation module. Specially, the motif-aware module helps calculate the correlation between nodes based on network motifs, and assigns weights to different adjacent nodes. The graph convolutional autoencoder module learns representations of nodes combining the high-order correlation obtained by the motif-aware layer. The tenuous subset generation part aims to get a node set that meets the conditions by leveraging the node representations.

\subsection{Motif-aware module}

Network motifs are building blocks in complex networks and are essential to the understanding of higher-order structural information in networks. This module helps calculate the higher-order correlation between nodes based on 3-node motifs and assign weights to different connected nodes. 

\textbf{Definition 3.} 3-node motif

A 3-node motif $M$ is defined on three nodes by a tuple $(A, S)$, where $A$ is the adjacency matrix, and $\mathrm{S}=\left\{S_{1}, S_{2}, \ldots\right\}$, where $S_{i}$ is the set of three nodes that participates in a triangle.
	
$S$ denotes a subset of $s$ nodes for defining the motif-based co-occurrence matrix. In other words, two connected nodes $u$ and $v$ will be regarded as co-occurring in a given motif only when they both belong to the same triangle.

We use the motif to capture the higher-order relationships between nodes in a graph when learning the network representations. Following the above definition of the 3-node motif, when given a motif set $(A, S)$, we use the co-occurrence of two nodes to capture the corresponding higher-order relationships. Assuming that the co-occurrence matrix constructed based on the 3-node motif is $M$, then the value of co-occurrence matrix is formed by node $i$ and node $j$. The formula is as follows:

\begin{equation}
M_{i j}=\sum_{k}(i, j) \in E \cap(i, j) \subseteq S_{k}
\end{equation}
\begin{figure}[htbp]
	\centerline{\includegraphics[scale=0.65]{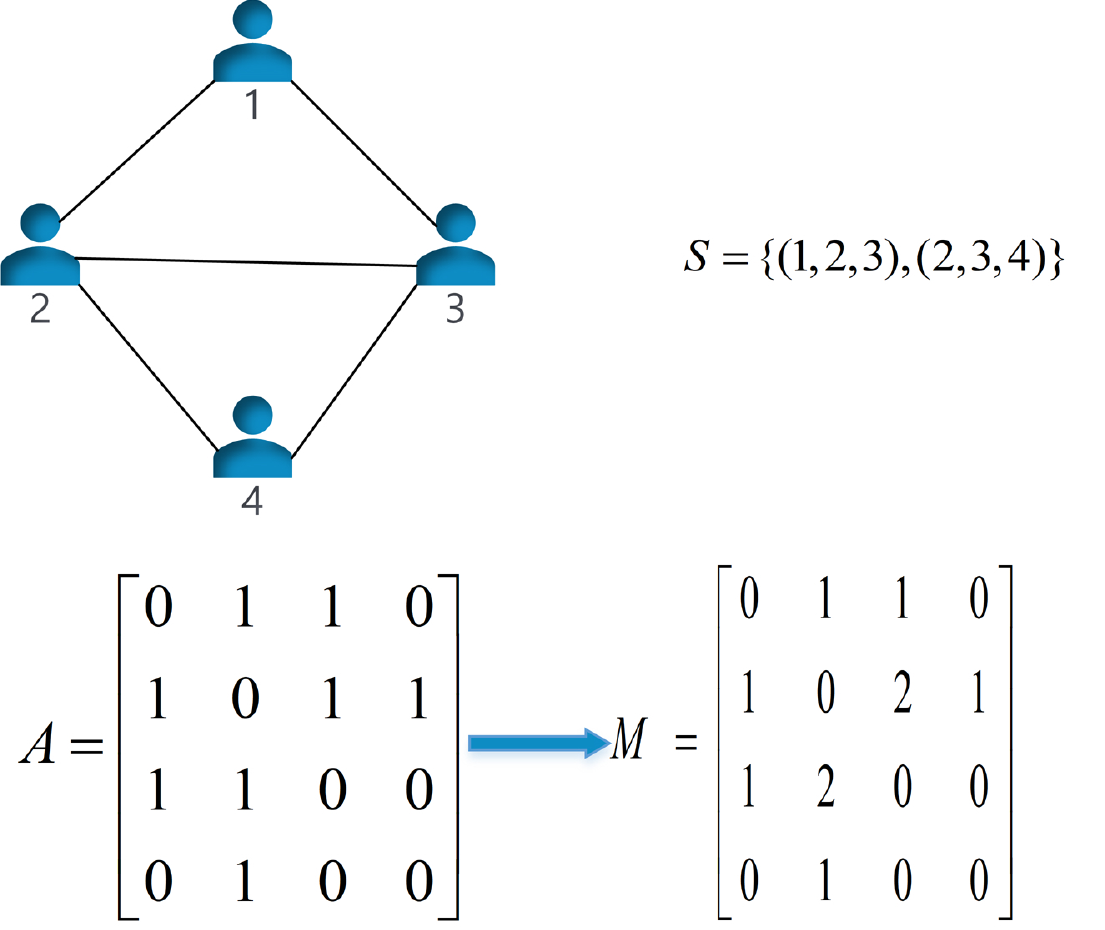}}  
	\caption{A toy example of 3-node motif-based adjacency matrix}
	\label{fig:motif}
\end{figure}
As shown in Fig. \ref{fig:motif}, there are four nodes in the network. The adjacency matrix is represented as $A$. $S$ is represented as $S_{1} \in\{1,2,3\},S_{2} \in\{2,3,4\}$, because node sets $\{1,2,3\}$ and $\{2,3,4\}$ form a 3-node motif, respectively. Nodes 2 and 3 appear in both $S_{1}$ and $S_{2}$ and participate together to form two 3-node motifs, thereby the value of $M_{2 3}$ is 2. And the number of 3-node motifs formed by node 1 and node 2 is only one, so the value of $M_{1 2}$ is 1.

Then we normalize the weight coefficient as follows:
\begin{equation}
\alpha_{i j}=\sigma_{j}\left(m_{i j}\right)+I=\frac{\exp \left(m_{i j}\right)}{\sum_{k \in \mathcal{N}_{i}} \exp \left(m_{i k}\right)}+I
\end{equation}
where $\sigma(\cdot)$ is the softmax function, $\mathcal{N}_{i}$ are the neighbors of node $i$, $I$ is the identity matrix with the same size as $M$.

\subsection{Graph convolutional autoencoder module}

The graph convolutional autoencoder aims to obtain a low-dimensional representation vector of nodes by fusing the adjacency matrix and feature matrix. It consists of two parts: encoder and decoder. 
We adopt GCN as the encoder. And the encoder can be represented as:
\begin{equation}
Z^{(l+1)}=f\left(Z^{(l)}, A \mid W^{(l)}\right)
\end{equation}
where $W^{(l)}$ is the weight matrix of the $l$-th layer which needs to be learned. $Z^{(l)}$ and $Z^{(l+1)}$ are the input and output of the $l$-th layer, respectively.

The nonlinear function $f(\cdot)$ is defined as:
\begin{equation}
f\left(Z^{(l)}, A \mid W^{(l)}\right)=\phi\left(D^{\prime\left(-\frac{1}{2}\right)} A D^{\prime\left(-\frac{1}{2}\right)} Z^{(l)} W^{(l)}\right)
\end{equation}
where $A^{\prime}=A+I$(we assume diagonal elements set to 1, i.e. every node is connected to itself), $D^{\prime}$ is the diagonal degree matrix, $D_{i i}^{\prime}=\sum_{j} A_{i j}$, $\phi$ is the activation function. In this paper, the encoder is constructed with a two-layer GCN.Each layer is represented as follows:
\begin{align}
Z^{(1)}&=f_{\text{relu}}(X,A|W^{(0)}) \\
Z^{(2)}&=f_{\text{linear}}(Z^{(1)},A|W^{(1)})
\end{align}
where $W^{(0)}$ and $W^{(1)}$ are the weight matrixes that need to be learned and they are trained using gradient descent. Here, for featureless, we simply drop the dependence on $X$ and replace $X$ with the identity matrix in the GCN. 

In general, the encoder encodes both structure information and node features into the representation $Z=\phi(Z \mid X, A)=Z^{2}$.

The decoder mainly uses the latent representation vector generated by encoder to reconstruct the graph structure. The reconstructed graph structure can be used to predict whether there is an edge between two nodes. Specifically, let the reconstructed adjacency matrix be $A^{\prime}$, the decoder is defined as an inner product between the representation vectors of nodes:
\begin{align}
\begin{split}
	p(A'|Z)=\prod_{i=1}^{N}\prod_{j=1}^{N}p(A'_{ij}|\vec{z}_i,\vec{z}_j), \\ 
	\text{with} \; p(A'_{ij}=1|\vec{z}_i,\vec{z}_j)=\sigma(\vec{z}_i^T\vec{z}_j)
\end{split}
\end{align}
where $\sigma(\cdot)$ is the sigmoid function.

The loss of graph autoencoder is defined as the reconstruction error between $A$ and $A^{\prime}$:
\begin{equation}
\mathcal{L}=\mathbb{E}_{\phi(z \mid X, A)}\left[\log p(A'|Z))\right]
\end{equation}

Considering the weights coefficient $a_{i j}$ calculated by motif-aware module, the final low-dimensional representation vector of nodes can be represented as follows:
\begin{equation}
Z^{(l+1)}=\sum_{j \in \mathcal{N}_{i}} a_{i j} Z^{(l)}
\end{equation}

To reduce the required memory, we use the sparse representation of adjacency matrix $A$ as the initial graph representation. So, the memory consumption is $O(|E|)$.

\subsection{Tenuous subset generation module}

This module leverages obtained nodes representation and design algorithms in vector space. In order to solve the problem of tenuous subgraph finding, we perform a heuristic algorithm based on local density expansion in low-dimensional vector space. 

\textbf{Definition 4.} $\varepsilon-$ pair

Given the representation vectors of two nodes, we call it a $\varepsilon-$ pair if their distance is not greater than $\varepsilon$. Here, we use Euclidean distance to measure the distance of two nodes in vector space.
\begin{equation}
\operatorname{dis}\left(x_{i}, x_{j}\right)=\sqrt{\sum_{t=1}^{h}\left(x_{i t}-x_{j t}\right)^{2}}
\end{equation}

With the definition of $\varepsilon-$-pair, we can measure the tenuity of a given vector set. Assuming that the number of $\varepsilon-$ pairs in tenuous subset $T$ is $r$, the smaller $r$, the more tenuous the set $T$.

\textbf{Definition 5.} $\varepsilon-$ neighbor($N_{\varepsilon}$)

Different from the definition of neighbors in the graph as the nodes with edges between them, the vector space has no explicit edge information. However, the distance between any nodes can be easily calculated. These distances reflect the relationship between nodes in the original graph. If the distance between two nodes is small, then we consider them as neighbors. Accordingly, for any node $u \in V$, given a distance parameter $\varepsilon$, the $\varepsilon-$neighbor of $u$ is defined as follows:
\begin{equation}
N_{\varepsilon}(u)=\{v \in V \mid \operatorname{dis}(u, v) \leq \varepsilon\}
\end{equation}

\begin{figure}[htbp]
	\centerline{\includegraphics[scale=0.85]{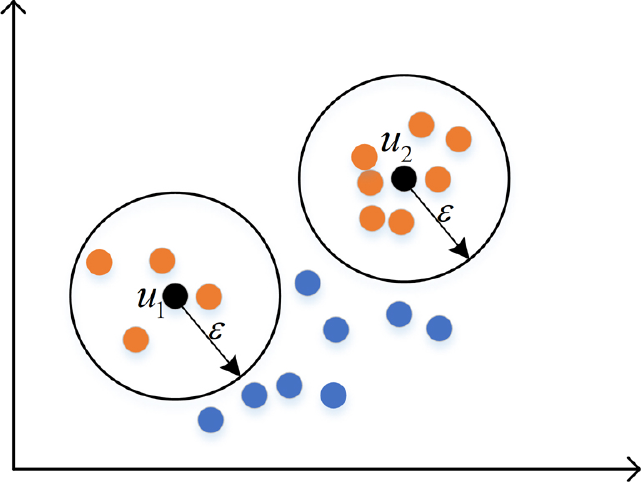}} 
	\caption{An example of $\varepsilon-$neighbor}
	\label{fig:neighbor}
\end{figure}

As shown in Fig \ref{fig:neighbor}, we draw circles with radius $\varepsilon$ centered on node $u_{1}$ and $u_{2}$, respectively, and we use the big circle to denote the neighborhood boundary. Nodes in orange within the circle are the $\varepsilon-$neighbors of $u_{1}$ and $u_{2}$, respectively.

\textbf{Definition 6.} density ($\rho(u)$)

Intuitively, the number of neighbors represents the degree of interaction between the node and the other nodes. Therefore, the tenuity of a node can be simply expressed as the number of neighbors of it. 
\begin{equation}
\rho(u)=\frac{\left|N_{\varepsilon}(u)\right| \cap V}{|V|}
\end{equation}

Take Fig. \ref{fig:neighbor} as an example, we show how to calculate density $\rho$. We can see from Fig. \ref{fig:neighbor}, node $u_{1}$ has four $\varepsilon-$neighbors and there are 21 nodes in total, thus $\rho\left(u_{1}\right)=\frac{4}{21}$. While node $u_{2}$ has seven $\varepsilon-$neighbos, thus $\rho\left(u_{2}\right)=\frac{7}{21}$.

Next, we propose a heuristic algorithm based on local density in low-dimensional vector space. Empirically, a greater distance constraint $\epsilon$ means a smaller subgraph. In other words, the size of the obtained tenuous subset is negatively related to the distance constraint. If the size of obtained subgraph does not meet the constraint, we reduce the distance constraint $\varepsilon$ to a smaller $\varepsilon^{\prime}$. Therefore, we can perform the binary search on the distance constraint for efficient calculation.

The algorithm first calculates the number of $\varepsilon-$neighbor nodes and neighbors of all nodes; then selects an optimal node, adds it to the tenuous subset and deletes all nodes in its $\varepsilon-$neighbor; finally, for the rest candidate nodes, iteratively selects the optimal node which meets the conditions and adds it to the tenuous subset. The algorithm always selects the optimal node in current state when adding nodes to the tenuous subset.

There are two strategies to select optimal nodes which can be seen as a trade-off between search efficiency and subgraph size. 

The former prioritizes the selection of the nodes with largest $\rho$ to join the tenuous subset. Because in this way, the number of remaining optional nodes in the next iteration will be reduced, that is, the number of iterations will be reduced. Thus, this method has high search efficiency. The latter prioritizes the selection of the nodes with smallest $\rho$ to join the tenuous subset. After we select a node $u$ with smallest $\rho$, we need to delete nodes which are the neighbors of $u$. So, we will obtain a subgraph with more nodes which can easily meet the size constraint. Besides, the running time is slightly longer than that of former. 

\begin{algorithm}
	\caption{tenuous subset generation with lower bound on the size }  
	\KwIn{$V=\left\{v_{1}, v_{2}, \cdots, v_{n}\right\},$ node vector set $R^{h}$, size constraint $k$ distance parameter $d$. }
	\KwOut{tenuous subset $T$.}  
	unvisited set $U \leftarrow V$
	
	tenuous subset $T \leftarrow \emptyset$
	
	\For{$v_{i} \in V$}
	{
		find out the $\varepsilon$ neighbor of node $v_{i}: N_{\varepsilon}\left(v_{i}\right),$ and calculate the number of $\varepsilon$-neighbors: $\left|N_{\varepsilon}\left(v_{i}\right)\right|$.
	}
	
	\While{$|T|$<$k$}  
	{
		
		\While {$U \neq \emptyset$}
		
		{
			$v_{t} \leftarrow$ the node with biggest density
			
			$T \leftarrow T \cup\left\{v_{t}\right\}$
			
			delete node $v_{t}$ and all its neighbors
		}
	
		$d \leftarrow d^{\prime}$
		
		return $T$
	}
\label{tab:pseudo}
\end{algorithm} 

Here, we give the pseudo by prioritizing the selection of the nodes with largest density to join the tenuous subset in Table \ref{tab:pseudo}.

\textbf{Time Complexity}

When calculating neighbors, by adopting the kd-tree data index structure, the time complexity is $O(NlogN)$. When node is added to the tenuous subset, the linear search method is used to find the node with the smallest number of neighbors, and the time complexity is $O(N)$. The time complexity of deleting nodes in the neighbors of a given node is $O(p)$, where $p$ is the average number of neighbors of each node. In summary, the total time complexity of the tenuous subset discovering process is $O(l(N+p))$, where $l$ is the size of obtained tenuous subsets, usually $l \ll N $. It can be approximated that the time complexity of the tenuous subset generation is nearly linear. Note that the process of calculating neighbors can be done efficiently offline. Therefore, the time complexity of GNNM algorithm is approximately $O(l(N+p))$.

\section{Experiments and Results}

In this section, we first describe the datasets and then evaluate the efficiency and effectiveness of GNNM with two baselines TERA\cite{Shen2020}  and WK\cite{Li2018}. All algorithms are coded in Python, and all the experiments are implemented on a computer with a 3.20GHz CPU and 64GB memory.

\subsection{Experimental Setup}

\subsubsection{Datasets}

In the experimental part, we use several real-world datasets and synthetic datasets. The detailed information of these data sets are summarized in the following tables, in which $|V|$ and $|E|$ represent the number of nodes and edges, respectively.

Table \ref{tab:real_world_datasets} depicts three real-world datasets. The details of these datasets are described as follows.

\begin{table}[]
	\centering
	\caption{Real-world datasets}
	\setlength{\tabcolsep}{9mm}{ 
	\begin{tabular}{ccccc}
		\hline
		Dataset  & $|V|$   & $|E|$   & Average Degree & Max Degree     \\ 
		\hline
		Hvr      & 304     & 3,263   & 21.467         & 61             \\ 
		Cora     & 2,708   & 5,429   & 4.009          & 169            \\ 
		Citeseer & 3,264   & 4,532   & 2.777          & 99             \\  
		\hline
	\end{tabular}}
	\label{tab:real_world_datasets}
\end{table}

Hvr\cite{Sun}: The Hvr dataset is a gene network consisting of several networks of highly recombinant malaria parasite genes. It contains 304 nodes and 3,263 edges.

Cora\cite{Yang}: The Cora dataset is a paper citation network which contains 2,708 nodes and 5,429 edges. The nodes represent the papers, and the edges represent the citation relationships between papers. 

Citeseer\cite{Lu}: The Citeseer dataset is a citation network which contains 3,264 nodes and 4,532 edges. The nodes represent the publications, and the edges represent the links between publications.

\begin{table}[]
	\centering
	\caption{Synthetic datasets for efficiency analysis}\label{tab:efficiency analysis}
	\setlength\tabcolsep{3.5pt}{
	
	\setlength{\tabcolsep}{7mm}{ 
	\begin{tabular}{ccccc}
		\hline
		Dataset   & $|V|$ & $|E|$ & Average Degree &Max Degree    \\
		\hline 
		Synthetic\_1000  	& 1,000  & 1,373   & 2.746   &5      \\
		Synthetic\_5000 	& 5,000  & 6,875   & 2.750   &5      \\
		Synthetic\_10000  	&10,000  & 16,757  & 3.351   &5      \\
		Synthetic\_50000	&50,000  & 69,104  & 2.764   &5      \\
		Synthetic\_100000	&100,000 & 168,026 & 3.360   &5      \\ 
           \hline
	\end{tabular}}
	}
\end{table}

\begin{table}
\centering
\caption{Synthetic datasets for effectiveness analysis}

\setlength{\tabcolsep}{8mm}{ 
\begin{tabular}{ccccc} 
\hline
Dataset & $|V|$    & $|E|$      & Average Degree & Max Degree  \\ 
\hline
\multirow{7}{*}{Synthetic\_1000} 
& 1,000 & 1,373   & 2.746          & 5           \\
& 1,000 & 3,066   & 6.132          & 8           \\
& 1,000 & 4,401   & 8.802          & 10          \\
& 1,000 & 5,144   & 10.288         & 12          \\
& 1,000 & 6,873   & 13.746         & 16          \\
& 1,000 & 7,421   & 14.842         & 19          \\
& 1,000 & 8,082   & 16.164         & 19          \\ 
\hline
\multirow{7}{*}{Synthetic\_2000} 
& 2,000 & 2,341   & 2.341          & 3           \\
& 2,000 & 5,783   & 5.783          & 6           \\
& 2,000 & 8,856   & 8.856          & 10          \\
& 2,000 & 10,259  & 10.259         & 12          \\
& 2,000 & 12,517  & 12.517         & 14          \\
& 2,000 & 14,906  & 14.906         & 16          \\
& 2,000 & 15,741  & 15.741         & 18          \\ 
\hline
\multirow{7}{*}{Synthetic\_5000} 
& 5,000 & 6,875   & 2.750          & 5           \\
& 5,000 & 15,235  & 6.094          & 8           \\
& 5,000 & 22,020  & 8.808          & 10          \\
& 5,000 & 25,595  & 10.238         & 12          \\
& 5,000 & 30,695  & 12.278         & 14          \\
& 5,000 & 36,870  & 14.748         & 17          \\
& 5,000 & 41,840  & 16.736         & 19          \\
\hline
\end{tabular}}
\label{tab:effectiveness analysis}
\end{table}

We also generate several synthetic datasets. The details of these datasets are listed in Table \ref{tab:efficiency analysis} and Table \ref{tab:effectiveness analysis}. The datasets in Table \ref{tab:efficiency analysis} are used for efficiency analysis, while the datasets in Table \ref{tab:effectiveness analysis} are used for effectiveness analysis.

\subsubsection{Compared Algorithms and Evaluation Metrics}

In order to evaluate the performance of our proposed algorithm GNNM, we compare two existing baseline algorithms WK and TERA with it in this section. 
\begin{itemize}
	\item TERA: TERA is proposed for MKTG problem which aims to minimize the number of $k$-triangles. 
	\item WK: WK is an improvement of TERA which aims to minimize the number of $k$-lines. 
\end{itemize}

We use $k$-line, $k$-triangle and $PF$ to evaluate the effectiveness of our proposed algorithm GNNM.
\begin{equation}
k-\operatorname{line}=\sum_{\forall i \in T} \sum_{\forall j \in T} d_{G}(i, j) \leq k
\end{equation}
\begin{equation}
k-\text { triangle }=\sum_{\forall(u, v, w) \in T} d_{G}(u, v) \leq k \text { and } \\
d_{G}(u, w) \leq k \text { and } d_{G}(v, w) \leq k
\end{equation}
where $d_{G}(u, v)$ is the shortest distance between node $u$ and $v$ in graph $G$.

\begin{table}
	\centering
	\caption{Results of different algorithms on three datasets}
	
	\setlength{\tabcolsep}{9mm}{
	\begin{tabular}{c|cccc} 
		\hline
		Dataset & Algorithm & 1-line & 1-triangle & $|PF|$  \\ 
           \hline
		\multirow{3}{*}{Synthetic\_1000} 
		& WK        & 0      & 0          & 139             \\
		& TERA      & 46     & 0          & 157             \\
		& GNNM      & 32     & 0          & \textbf{32}     \\ 
		\hline
		\multirow{3}{*}{Synthetic\_2000} 
		& WK        & 0      & 0          & 312             \\
		& TERA      & 71     & 0          & 381             \\
		& GNNM      & 44     & 0          & \textbf{44}     \\ 
		\hline
		\multirow{3}{*}{Synthetic\_5000} 
		& WK        & 0      & 0          & 277             \\
		& TERA      & 46     & 0          & 130             \\
		& GNNM      & 75     & 0          & \textbf{75}     \\
		\hline
	\end{tabular}}
	\label{tab:4}
\end{table}

Table \ref{tab:4} shows the results of the three algorithms on the Synthetic\_1000, Synthetic\_2000, and Synthetic\_5000 datasets. From this table,  we have the following observations.

First, the number of $|PF|$ introduced by GNNM is equal to the number of 1-lines indicates that nodes in the obtained subgraph have no potential friendship. In other words, these nodes have few interactions and weak relationships. 

Second, we find that the number of 1-lines obtained by TERA is the largest. The reason is that TERA only ensures that the obtained subgraph does not contain 1-triangles, so it could get a subgraph with no triangles but many connected edges which demonstrates the limitation of $k$-triangle. 
The number of $|PF|$ obtained by WK is the largest. The reason is that WK does not consider potential friendship between nodes, so it can get a subgraph with no 1-line, but more $|PF|$, which reveals the limitation of $k$-line. In general, the results demonstrate the effectiveness of GNNM and the necessity of $PF$.

\subsubsection{Running Time}

\begin{figure}[htbp]
	\centerline{\includegraphics[scale=0.45]{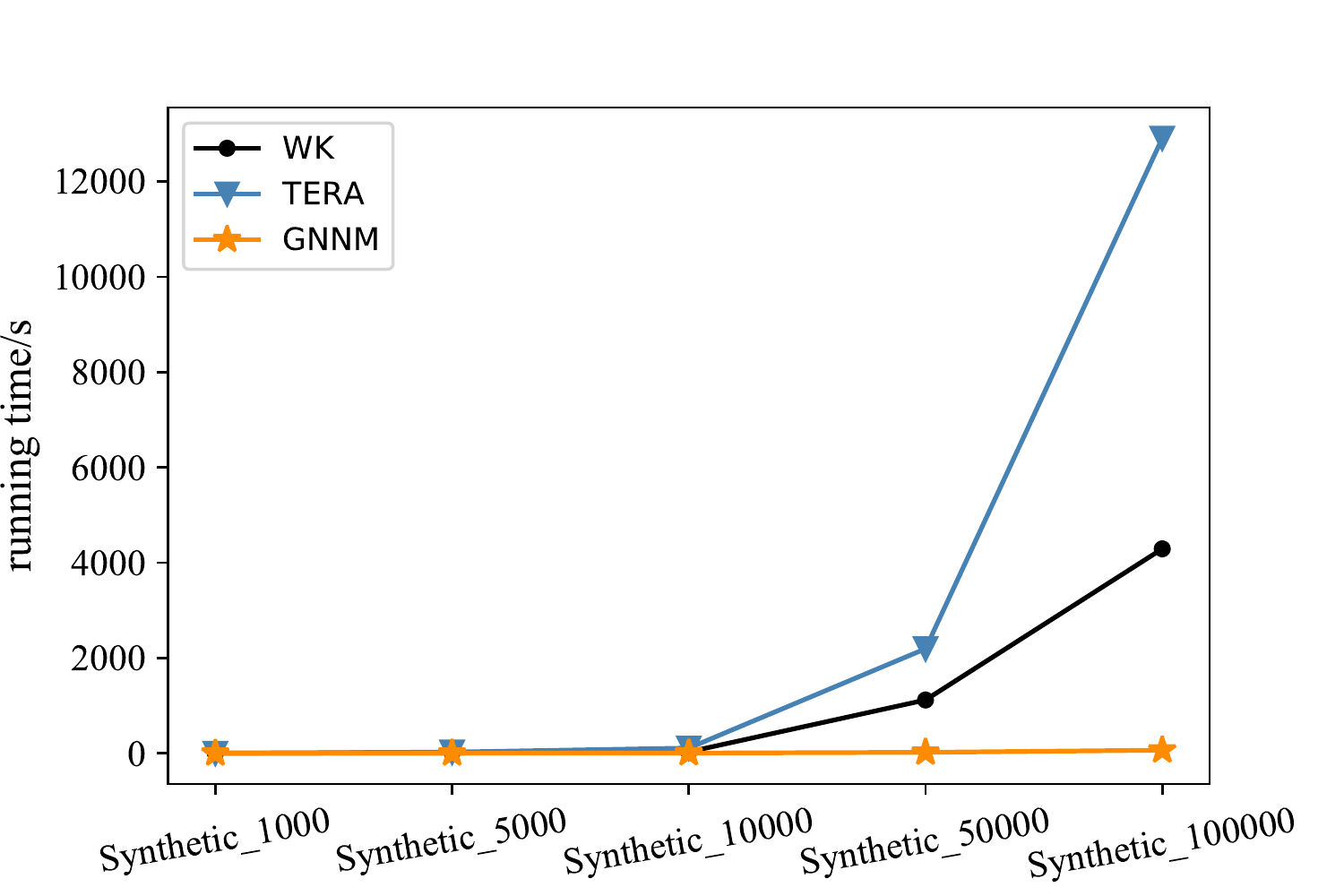}} 
	\caption{Running time of different algorithms on synthetic datasets}
	\label{fig:running time}
\end{figure}

In this section, we mainly compare the time efficiency of the proposed algorithm GNNM with TERA and WK.
Fig. \ref{fig:running time} reveals the running time of different algorithms on several synthetic datasets. From Fig. \ref{fig:running time}, we can observe that compared with the other two algorithms, GNNM has the highest time efficiency, while TERA algorithm has the lowest time efficiency. Because TERA needs to calculate the number of $k$-triangles that each node participates in, which is time-consuming. On the contrary, by mapping the original graph to the low-dimensional vector space, the time complexity of distance calculation in the vector space is greatly reduced compared with the shortest path calculation in the graph. Accordingly, our method presents superior performance.

\subsection{Parameter Sensitivity}

\begin{figure}[htbp]
	\centering
	\subfigure[the relationship between $\varepsilon$ and $|PF|$]{
		\label{fig:parameter sensitivity(a)}
			\includegraphics[width=0.45\textwidth]{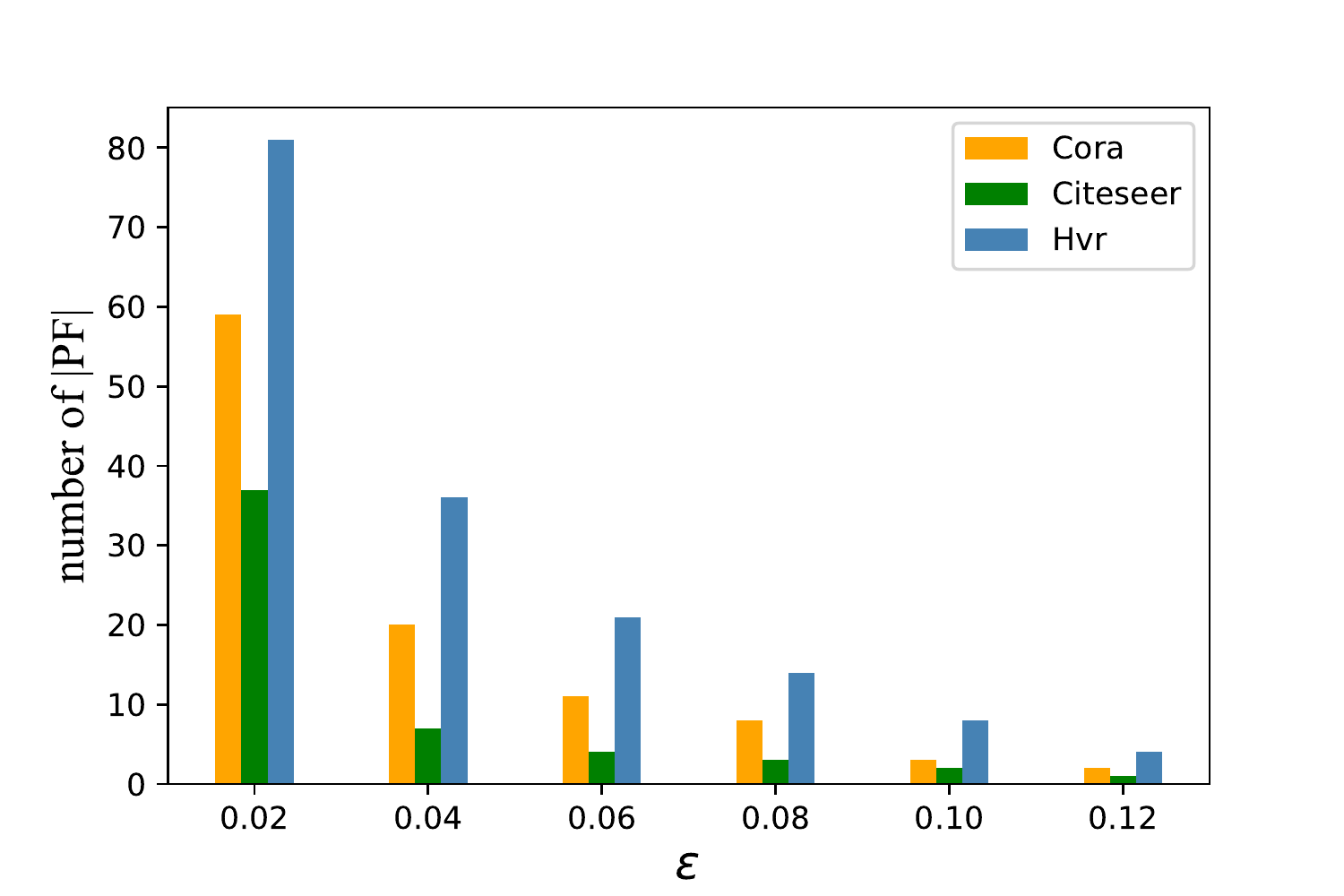}
	}
	\subfigure[the relationship between $\varepsilon$ and group size]{
		\label{fig:parameter sensitivity(b)}
			\includegraphics[width=0.45\textwidth]{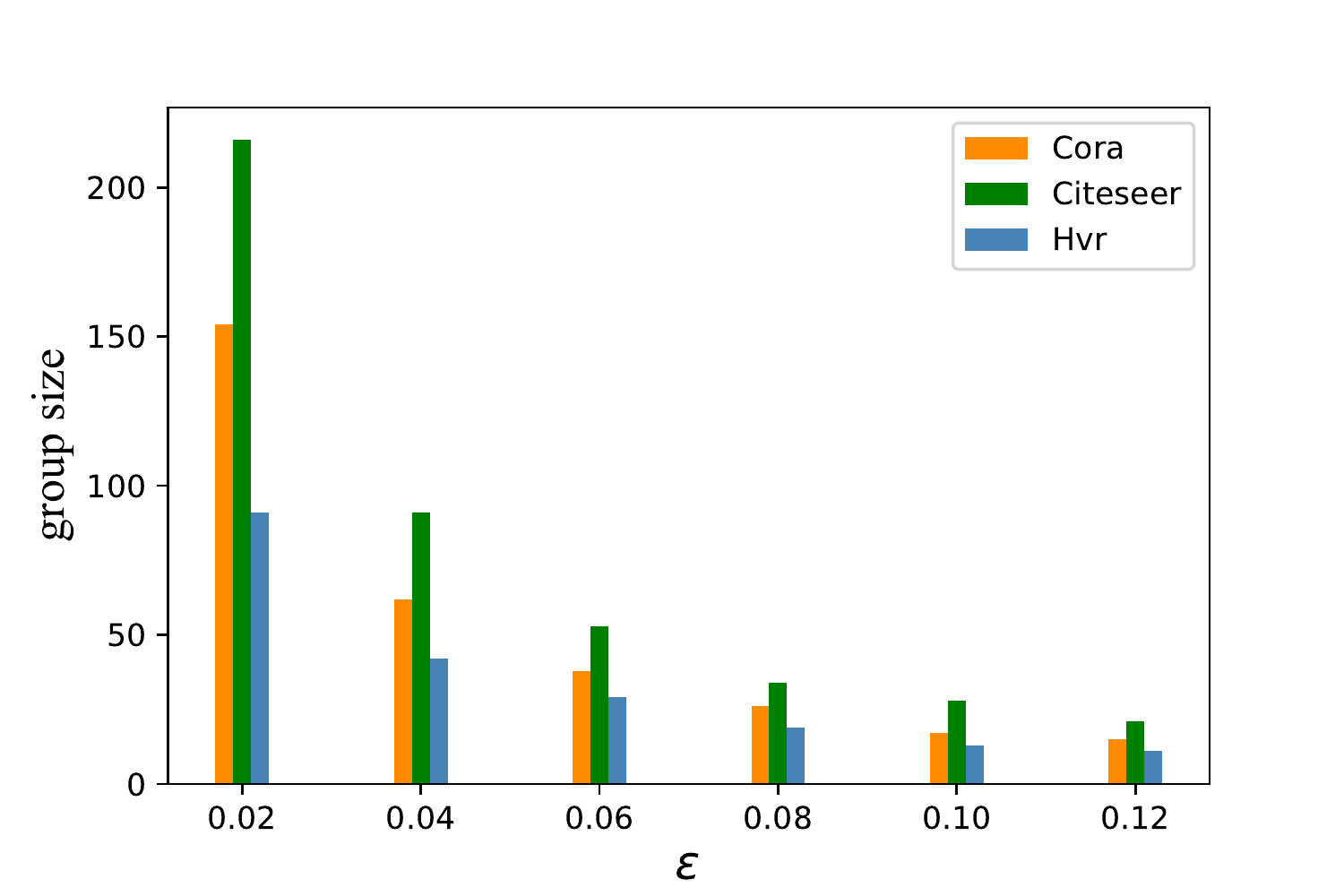}
	}
	\caption{Parameter sensitivity}
	\label{fig:parameter sensitivity(}
\end{figure}

Fig. \ref{fig:parameter sensitivity(a)} shows the relationship between $\varepsilon$ and the number of $|PF|$. Fig. \ref{fig:parameter sensitivity(b)} shows the relationship between $\varepsilon$ and the obtained graph size. From Fig. \ref{fig:parameter sensitivity(a)}, we find that the number of $|PF|$ obtained by Hvr is the largest because it is denser than others. As shown in Fig. \ref{fig:parameter sensitivity(b)}, when $\varepsilon$ becomes large, small graphs are preferred because larger graphs obtained by dense graphs tend to incur much more potential friends. Moreover, Citeseer can find a larger graph without generating any potential friends because the average degree of Citeseer is small.

\subsection{Ablation Study}

In this section, we use three synthetic datasets to evaluate the efficiency and effectiveness of GNNM. We denote the variant of GNNM as Basic which is not based on motif.

\begin{figure}[htbp]
	\centering
	\subfigure[Synthetic\_1000 dataset]{
		\label{fig:degree(a)}
			\includegraphics[width=0.3\textwidth]{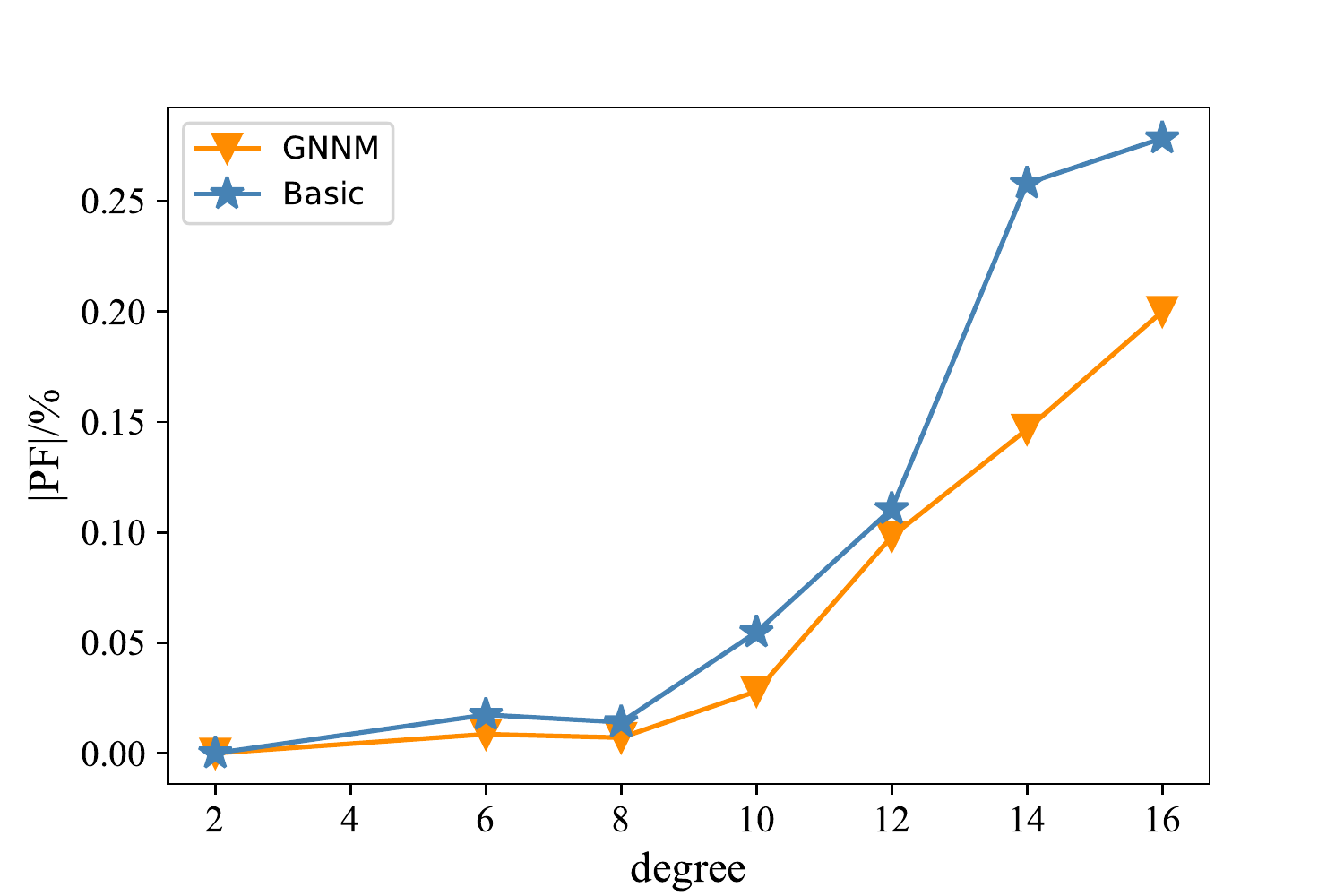}
	}
	\subfigure[Synthetic\_2000 dataset]{
		\label{fig:degree(b)}
			\includegraphics[width=0.3\textwidth]{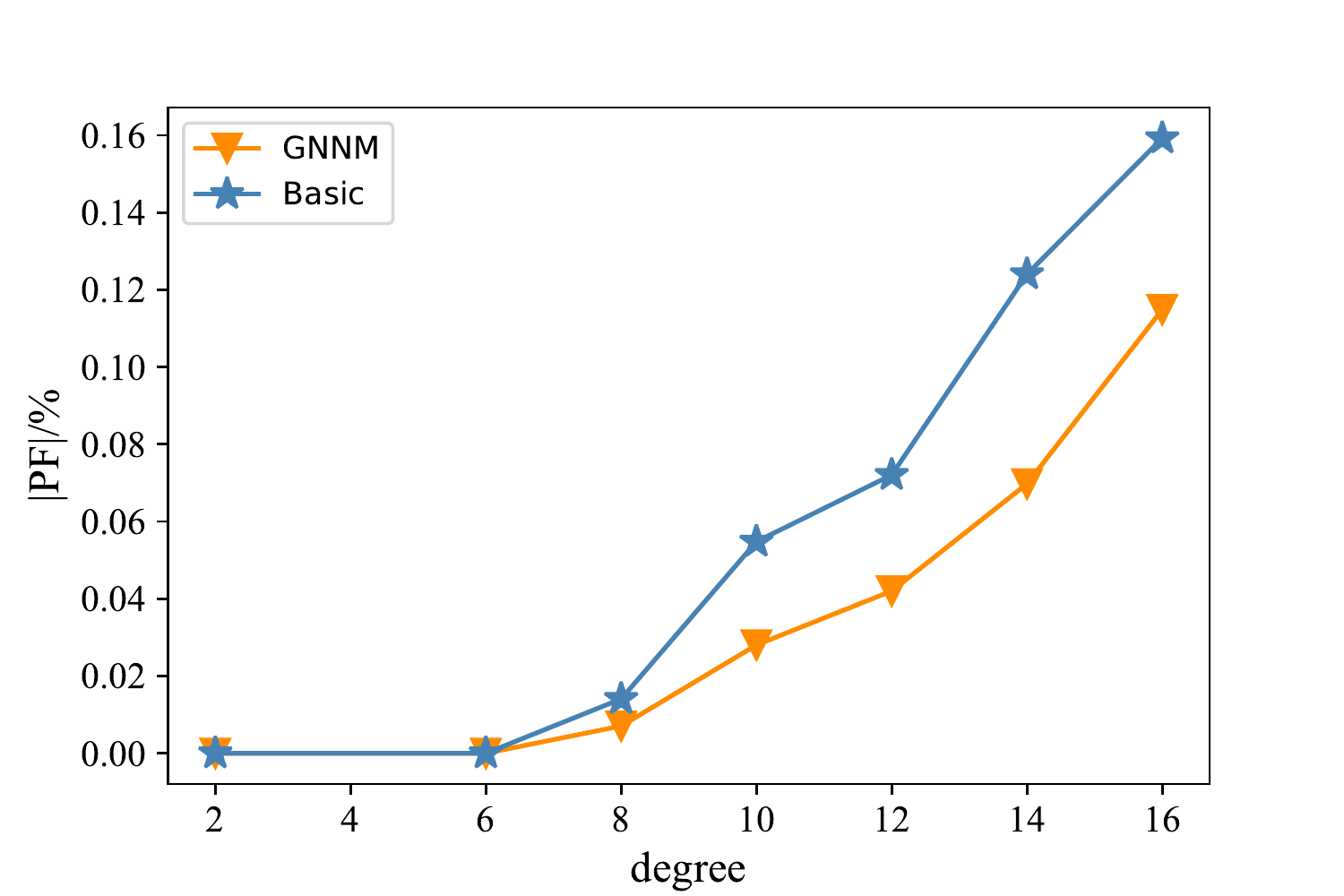}\
	}
	\subfigure[Synthetic\_5000 dataset]{
		\label{fig:degree(c)}
			\includegraphics[width=0.3\textwidth]{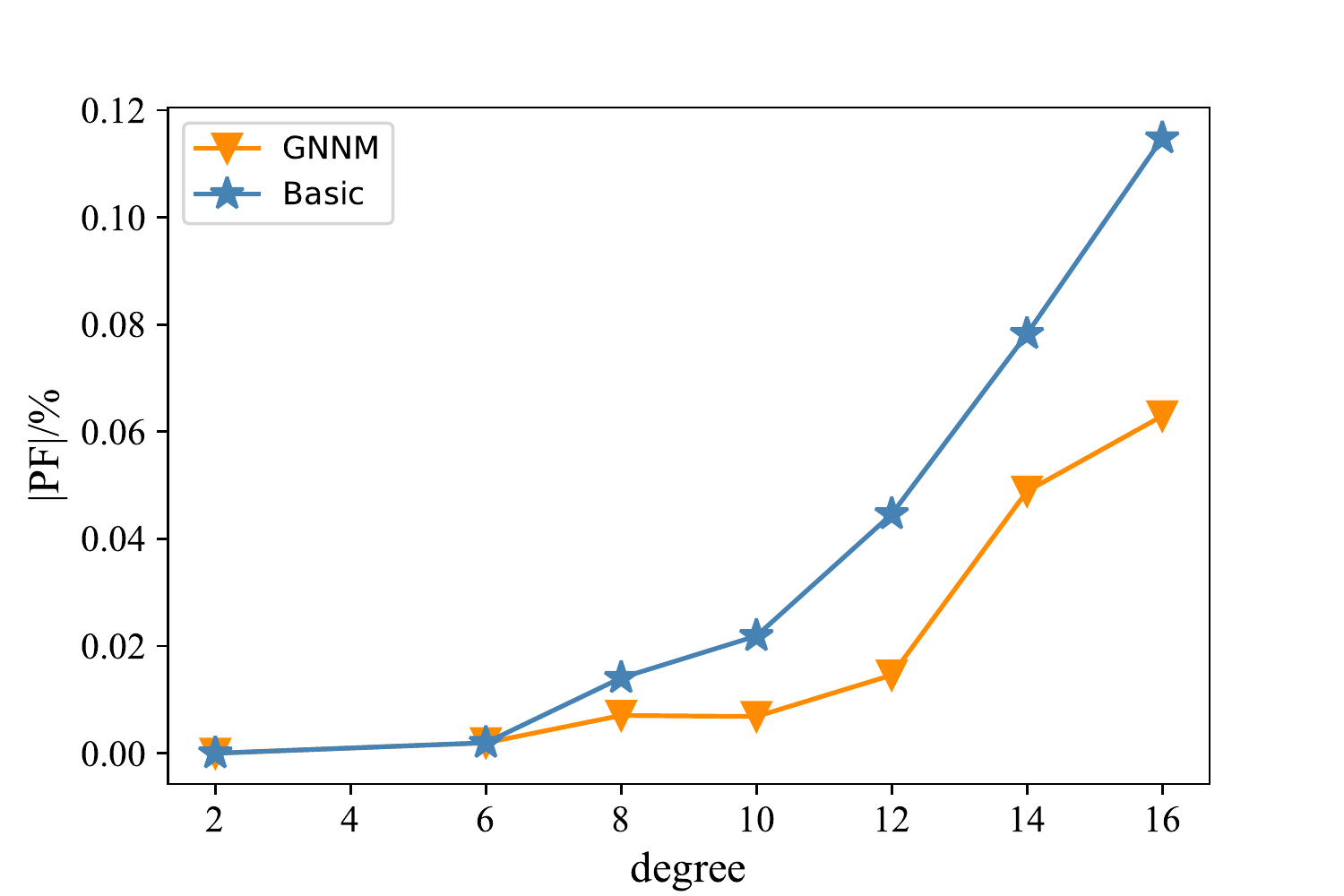}
	}
	\caption{The relationship between degree and $|PF|$}
	\label{fig:degree}
\end{figure}

Fig. \ref{fig:degree} shows the number of $|PF|$ in the obtained tenuous subgraph by Basic and GNNM on the three synthetic datasets, respectively. As shown in the figure, GNNM outperforms Basic. The number of $|PF|$ obtained by GNNM is less than that of Basic. Especially, with the increase of degree, the difference between the two algorithms becomes obvious. This is because the average degree of this network is relatively large, and there are more 3-node motifs in the network. GNNM can capture this higher-order structure by network representation based on the 3-node motif. Thereby, it can reduce the number of $|PF|$ in the obtained subgraph.

\begin{figure}[htbp]
	\centering
	\subfigure[Synthetic\_1000 dataset]{
			\label{ablation_1000_group}
			\includegraphics[width=0.3\textwidth]{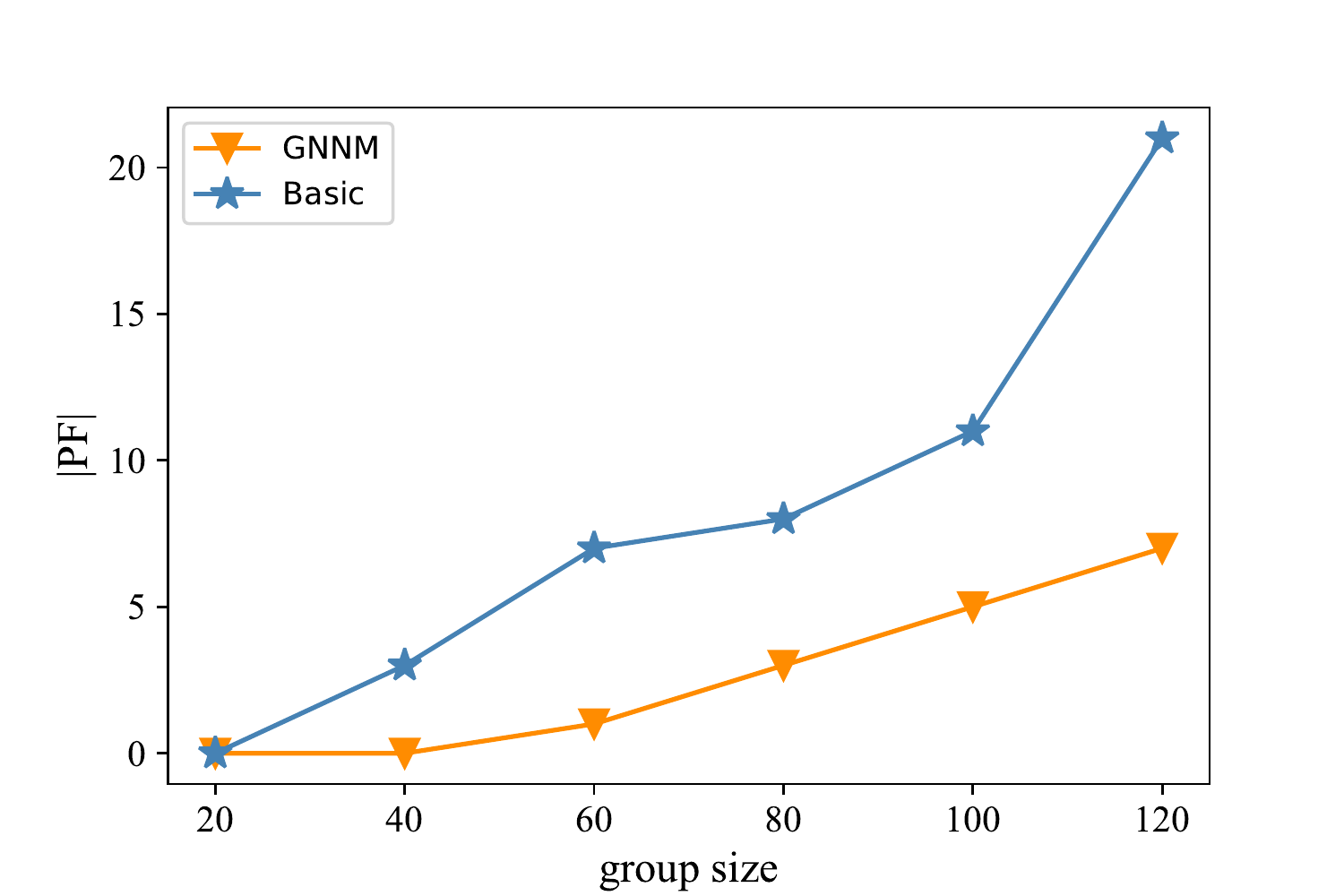}
			
	}
	\subfigure[Synthetic\_2000 dataset]{
			\label{ablation_2000_group}
			\includegraphics[width=0.3\textwidth]{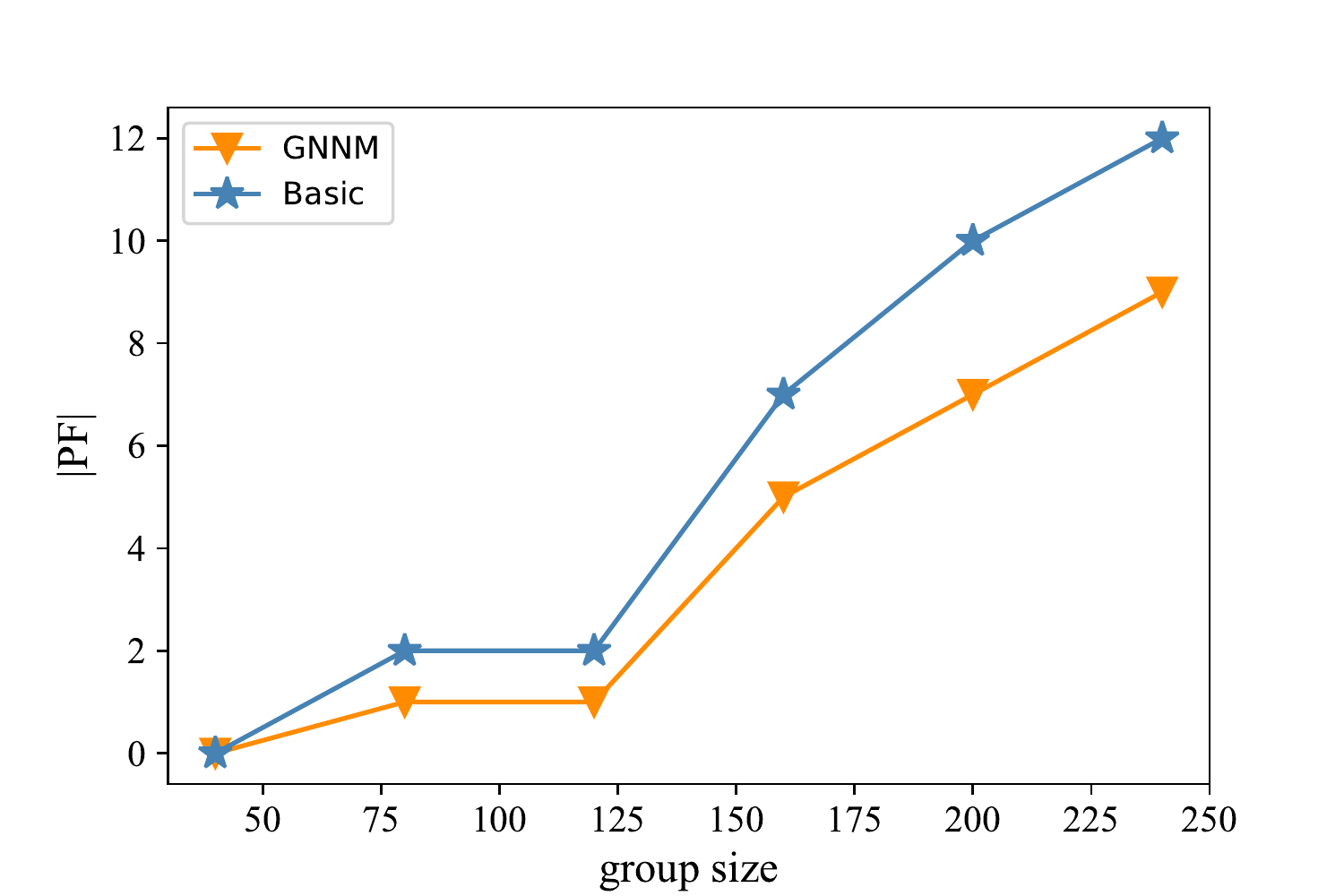}
			
	}%
	\subfigure[Synthetic\_5000 dataset]{
			\label{ablation_5000_group}
			\includegraphics[width=0.3\textwidth]{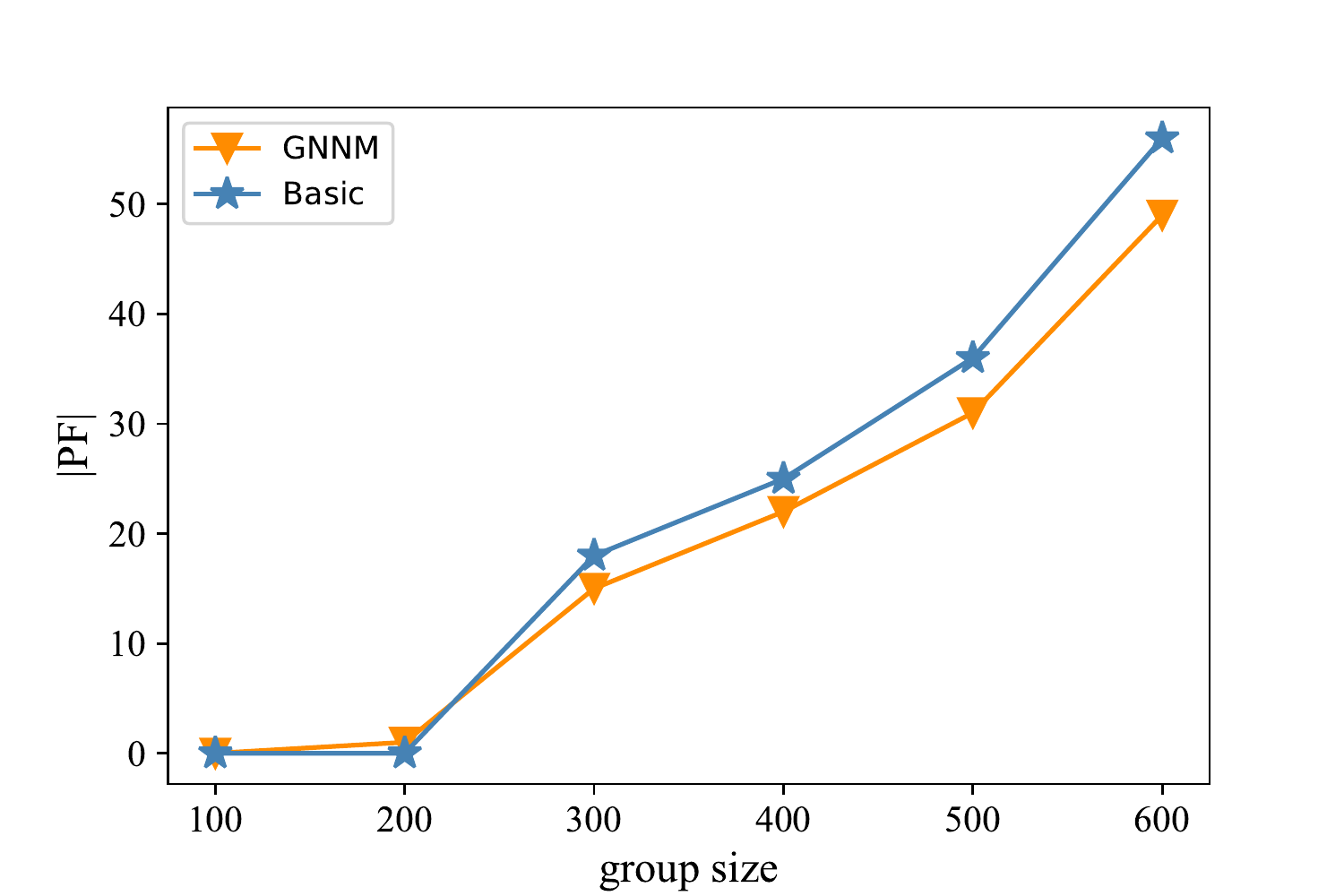}
	}%
	\caption{The relationship between group size and $|PF|$}
	\label{fig:group size}
\end{figure}

Fig. \ref{fig:group size} shows the relationship between the obtained subgraph size and $|PF|$. As shown in the figure, when graph size increases, the number of $|PF|$ also increases. Moreover, the number of $|PF|$ obtained by Basic is larger than that of GNNM. And the difference becomes bigger and bigger with the increase of graph size. The reason is that GNNM uses the motif-aware module to pay more attention to node pairs that participate in more 3-node motifs. Such node pairs are more similar with each other and should be avoided to be added to the final tenuous subgraph. Thus GNNM can reduce the number of the latent triangles in the otained tenuous subgraph.

\subsection{Case Study}

\begin{figure}[htbp]
	\centerline{\includegraphics[scale=0.3]{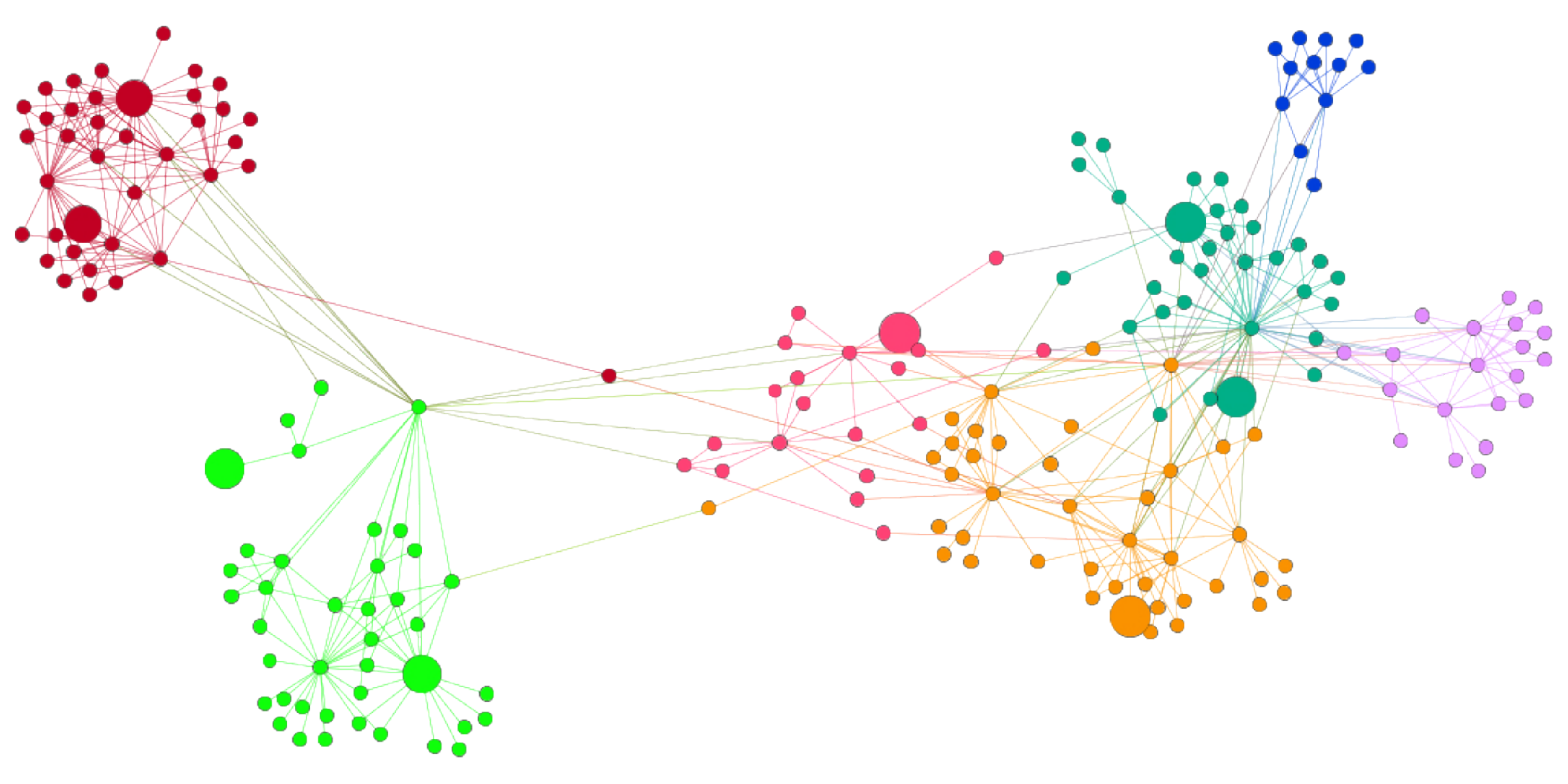}} 
	\caption{Tenuous subgraph found in Aminer network}
	\label{fig:9}
\end{figure}

In order to show the performance of GNNM, we conduct a case study on the Aminer dataset which was extracted and mined from academic social networks. We use this dataset to do reviewer selection and visualize the result. We select one of the subnet from Aminer and show the result in Fig. \ref{fig:9}. Larger nodes are the ones obtained by GNNM which form a tenuous subgraph whose names are Daniel C. McFarlane, Robert Gray, Seyed Amir Iranmanesh, Haifeng Chen, Nick Feamster, Isao Nagasawa, Ben Y. Zhao and Fabian E. Bustamante. Their research fileds contain network and system measurement and management. They work in different universities and have less intersections. We can notice from the figure that these nodes are evenly distributed. And we can also see that there are no edges in the tenuous subgraph and the relationship between nodes is very weak. Since we set size constraint $K$ to 8, we only obtain a subgraph with 8 nodes. If we increase the value of $ K$, then we can find a larger tenuous subgraph.

\section{Conclusion}

In this paper, we proposed a novel method for tenuous subgraph finding based on motif-based network representation. Considering the higher-order relationship between nodes based on network 3-node motif, we developed a framework that consists of three modules: motif-aware module, graph autoencoder with motif-aware module, and tenuous subset generation module. Extensive experiments are designed to demonstrate the efficiency and effectiveness of the motif-based network representation. The experimental results demonstrate the superiority of our model. Since we just simply use the node features information without guaranteeing the features of nodes in the tenuous subgraph should be closely related, we will study it in the future.

\bibliographystyle{unsrt}  
\bibliography{references}

\begin{thebibliography}{10}

\bibitem{YangWSDM2013}
Jaewon Yang and Jure Leskovec.
\newblock Overlapping community detection at scale: A nonnegative matrix
  factorization approach.
\newblock In {\em Proceedings of the Sixth ACM International Conference on Web
  Search and Data Mining}, WSDM '13, page 587–596, New York, NY, USA, 2013.
  Association for Computing Machinery.

\bibitem{Xie2011}
Xie Jierui, Stephen Kelley, and Boleslaw~K. Szymanski.
\newblock Overlapping community detection in networks: the state of the art and
  comparative study.
\newblock {\em CoRR}, abs/1110.5813, 2011.

\bibitem{huang2014}
Huang Xin, Cheng Hong, Qin Lu, Tian Wentao, and Yu~Jeffrey~Xu.
\newblock Querying k-truss community in large and dynamic graphs.
\newblock In {\em Proceedings of the 2014 ACM SIGMOD international conference
  on Management of data}, pages 1311--1322, 2014.

\bibitem{BothorelCMM15}
C{\'{e}}cile Bothorel, Juan~David Cruz, Matteo Magnani, and Barbora
  Micenkov{\'{a}}.
\newblock Clustering attributed graphs: models, measures and methods.
\newblock {\em CoRR}, abs/1501.01676, 2015.

\bibitem{sanei2018}
Seyed-Vahid Sanei-Mehri, Apurba Das, and Srikanta Tirthapura.
\newblock Enumerating top-k quasi-cliques.
\newblock In {\em 2018 IEEE International Conference on Big Data (Big Data)},
  pages 1107--1112. IEEE, 2018.

\bibitem{conte2018}
Alessio Conte, Tiziano~De Matteis, Daniele~De Sensi, Roberto Grossi, Andrea
  Marino, and Luca Versari.
\newblock D2k: scalable community detection in massive networks via
  small-diameter k-plexes.
\newblock In {\em Proceedings of the 24th ACM SIGKDD International Conference
  on Knowledge Discovery \& Data Mining}, pages 1272--1281, 2018.

\bibitem{Wen2019}
Dong Wen, Lu~Qin, Ying Zhang, Lijun Chang, and Ling Chen.
\newblock Enumerating k-vertex connected components in large graphs.
\newblock In {\em 2019 IEEE 35th International Conference on Data Engineering
  (ICDE)}, pages 52--63, Los Alamitos, CA, USA, apr 2019. IEEE Computer
  Society.

\bibitem{Liu2020}
Fanzhen Liu, Shan Xue, Jia Wu, Chuan Zhou, Wenbin Hu, C{\'{e}}cile Paris, Surya
  Nepal, Jian Yang, and Philip~S. Yu.
\newblock Deep learning for community detection: Progress, challenges and
  opportunities.
\newblock {\em CoRR}, abs/2005.08225, 2020.

\bibitem{U2001}
U.~Feige, D.~Peleg, and G.~Kortsarz.
\newblock The dense k-subgraph problem.
\newblock {\em Algorithmica}, 29(3):410–421, March 2001.

\bibitem{Shen2020}
Chih-Ya Shen, Shuai Hong-Han, Yang De-Nian, Lee Guang-Siang, Huang Liang-Hao,
  Lee Wang-Chien, and Chen Ming-Syan.
\newblock On extracting socially tenuous groups for online social networks with
  k-triangles.
\newblock {\em IEEE Transactions on Knowledge and Data Engineering}, pages
  1--1, 2020.

\bibitem{Li2018}
Li~Wentan.
\newblock Finding tenuous groups in social networks.
\newblock pages 284--291, 11 2018.

\bibitem{Milo824}
R.~Milo, S.~Shen-Orr, S.~Itzkovitz, N.~Kashtan, D.~Chklovskii, and U.~Alon.
\newblock Network motifs: Simple building blocks of complex networks.
\newblock {\em Science}, 298(5594):824--827, 2002.

\bibitem{Watrigant}
Marin Bougeret, Nicolas Bousquet, Rodolphe Giroudeau, and R{\'e}mi Watrigant.
\newblock Parameterized complexity of the sparsest k-subgraph problem in
  chordal graphs.
\newblock In Viliam Geffert, Bart Preneel, Branislav Rovan, J{\'u}lius
  {\v{S}}tuller, and A.~Min Tjoa, editors, {\em SOFSEM 2014: Theory and
  Practice of Computer Science}, pages 150--161, Cham, 2014. Springer
  International Publishing.

\bibitem{shen2015}
Chih-Ya Shen, Hong-Han Shuai, De-Nian Yang, Yi-Feng Lan, Wang-Chien Lee,
  Philip~S. Yu, and Ming-Syan Chen.
\newblock Forming online support groups for internet and behavior related
  addictions.
\newblock In {\em Proceedings of the 24th ACM International on Conference on
  Information and Knowledge Management}, CIKM '15, page 163–172, New York,
  NY, USA, 2015. Association for Computing Machinery.

\bibitem{hsu2018}
Bay-Yuan Hsu, Yi-Feng Lan, and Chih-Ya Shen.
\newblock On automatic formation of effective therapy groups in social
  networks.
\newblock {\em IEEE Transactions on Computational Social Systems},
  5(3):713--726, 2018.

\bibitem{li2020}
Yang Li, Heli Sun, Liang He, Jianbin Huang, Jiyin Chen, Hui He, and Xiaolin
  Jia.
\newblock Querying tenuous group in attributed networks.
\newblock {\em The Computer Journal}, 2020.

\bibitem{Cai2017}
HongYun Cai, Vincent~W. Zheng, and Kevin Chen-Chuan Chang.
\newblock A comprehensive survey of graph embedding: Problems, techniques, and
  applications.
\newblock {\em IEEE Transactions on Knowledge and Data Engineering},
  30(9):1616--1637, 2018.

\bibitem{Zhang2020}
Ziwei Zhang, Peng Cui, and Wenwu Zhu.
\newblock Deep learning on graphs: A survey, 2020.

\bibitem{deepwalk}
Bryan Perozzi, Rami Al-Rfou, and Steven Skiena.
\newblock Deepwalk: Online learning of social representations.
\newblock In {\em Proceedings of the 20th ACM SIGKDD international conference
  on Knowledge discovery and data mining}, pages 701--710. ACM, 2014.

\bibitem{node2vec}
Aditya Grover and Jure Leskovec.
\newblock node2vec: Scalable feature learning for networks.
\newblock In {\em Proceedings of the 22nd ACM SIGKDD international conference
  on Knowledge discovery and data mining}, pages 855--864. ACM, 2016.

\bibitem{line}
Jian Tang, Meng Qu, Mingzhe Wang, Ming Zhang, Jun Yan, and Qiaozhu Mei.
\newblock Line: Large-scale information network embedding.
\newblock In {\em Proceedings of the 24th International Conference on World
  Wide Web}, pages 1067--1077. International World Wide Web Conferences
  Steering Committee, 2015.

\bibitem{grarep}
Shaosheng Cao, Wei Lu, and Qiongkai Xu.
\newblock Grarep: Learning graph representations with global structural
  information.
\newblock In {\em Proceedings of the 24th ACM International on Conference on
  Information and Knowledge Management}, pages 891--900. ACM, 2015.

\bibitem{tadw}
Cheng Yang, Zhiyuan Liu, Deli Zhao, Maosong Sun, and Edward~Y. Chang.
\newblock Network representation learning with rich text information.
\newblock In {\em Proceedings of the 24th International Conference on
  Artificial Intelligence}, IJCAI'15, page 2111–2117. AAAI Press, 2015.

\bibitem{Wu2021}
Zonghan Wu, Shirui Pan, Fengwen Chen, Guodong Long, Chengqi Zhang, and
  Philip~S. Yu.
\newblock A comprehensive survey on graph neural networks.
\newblock {\em IEEE Transactions on Neural Networks and Learning Systems},
  32(1):4--24, 2021.

\bibitem{Zhou2021}
Jie Zhou, Ganqu Cui, Shengding Hu, Zhengyan Zhang, Cheng Yang, Zhiyuan Liu,
  Lifeng Wang, Changcheng Li, and Maosong Sun.
\newblock Graph neural networks: A review of methods and applications, 2021.

\bibitem{gae}
Thomas~N Kipf and Max Welling.
\newblock Variational graph auto-encoders.
\newblock {\em arXiv preprint arXiv:1611.07308}, 2016.

\bibitem{sdne}
Daixin Wang, Peng Cui, and Wenwu Zhu.
\newblock Structural deep network embedding.
\newblock In {\em Proceedings of the 22nd ACM SIGKDD international conference
  on Knowledge discovery and data mining}, pages 1225--1234. ACM, 2016.

\bibitem{dngr}
Shaosheng Cao, Wei Lu, and Qiongkai Xu.
\newblock Deep neural networks for learning graph representations.
\newblock In {\em Proceedings of the Thirtieth AAAI Conference on Artificial
  Intelligence}, AAAI'16, page 1145–1152. AAAI Press, 2016.

\bibitem{kipf2017}
Thomas~N. Kipf and Max Welling.
\newblock Semi-supervised classification with graph convolutional networks,
  2017.

\bibitem{gao2018}
Hongyang Gao, Zhengyang Wang, and Shuiwang Ji.
\newblock Large-scale learnable graph convolutional networks.
\newblock In {\em Proceedings of the 24th ACM SIGKDD International Conference
  on Knowledge Discovery \& Data Mining}, pages 1416--1424, 2018.

\bibitem{Ugander2013SubgraphFM}
J.~Ugander, L.~Backstrom, and J.~Kleinberg.
\newblock Subgraph frequencies: mapping the empirical and extremal geography of
  large graph collections.
\newblock {\em ArXiv}, abs/1304.1548, 2013.

\bibitem{Rahmtin2017}
Rahmtin Rotabi, Krishna Kamath, Jon~M. Kleinberg, and Aneesh Sharma.
\newblock Detecting strong ties using network motifs.
\newblock {\em CoRR}, abs/1702.07390, 2017.

\bibitem{bioinformatics}
Nataša Pržulj.
\newblock {Biological network comparison using graphlet degree distribution}.
\newblock {\em Bioinformatics}, 23:e177--e183, 2007.

\bibitem{Ahmed2015}
Ryan A.~Rossi Nesreen K.~Ahmed, Jennifer~Neville and Nick Duffield.
\newblock Efficient graphlet counting for large networks.
\newblock In {\em 2015 IEEE International Conference on Data Mining}, pages
  1--10, 2015.

\bibitem{HanS16}
Guyue Han and Harish Sethu.
\newblock Waddling random walk: Fast and accurate sampling of motif statistics
  in large graphs.
\newblock {\em CoRR}, abs/1605.09776, 2016.

\bibitem{Pinar2017}
Ali Pinar, C.~Seshadhri, and Vaidyanathan Vishal.
\newblock Escape: Efficiently counting all 5-vertex subgraphs.
\newblock In {\em Proceedings of the 26th International Conference on World
  Wide Web}, WWW '17, page 1431–1440, Republic and Canton of Geneva, CHE,
  2017. International World Wide Web Conferences Steering Committee.

\bibitem{Benson163}
Austin~R. Benson, David~F. Gleich, and Jure Leskovec.
\newblock Higher-order organization of complex networks.
\newblock {\em Science}, 353(6295):163--166, 2016.

\bibitem{Yin2017}
Hao Yin, Austin~R. Benson, Jure Leskovec, and David~F. Gleich.
\newblock Local higher-order graph clustering.
\newblock In {\em Proceedings of the 23rd ACM SIGKDD International Conference
  on Knowledge Discovery and Data Mining}, KDD '17, page 555–564, New York,
  NY, USA, 2017. Association for Computing Machinery.

\bibitem{Long2020}
Qingqing Long, Yilun Jin, Guojie Song, Yi~Li, and Wei Lin.
\newblock Graph structural-topic neural network.
\newblock In {\em Proceedings of the 26th ACM SIGKDD International Conference
  on Knowledge Discovery and Data Mining}, KDD '20, page 1065–1073, New York,
  NY, USA, 2020. Association for Computing Machinery.

\bibitem{Sun}
Heli Sun, Fang He, Jianbin Huang, Yizhou Sun, and Xiaolin Jia.
\newblock Network embedding for community detection in attributed networks.
\newblock {\em ACM Transactions on Knowledge Discovery from Data}, 14(3):1--25,
  2020.

\bibitem{Yang}
Z.~Yang, W.~W. Cohen, and R.~Salakhutdinov.
\newblock Revisiting semi-supervised learning with graph embeddings.
\newblock In {\em International Conference on International Conference on
  Machine Learning}, 2016.

\bibitem{Lu}
Q.~Lu and L.~Getoor.
\newblock Link-based classification.
\newblock {\em DBLP}, 2003.

\end{thebibliography}

\end{document}